\documentclass[fleqn,usenatbib]{mnras}

\usepackage{newtxtext,newtxmath}
\usepackage[T1]{fontenc}
\usepackage{ae,aecompl,times}

\usepackage{graphbox,graphicx}	% Including figure files
\usepackage{amsmath}	% Advanced maths commands
\usepackage{multirow}
\usepackage[normalem]{ulem}
\usepackage{xcolor}
\usepackage{bm}
\usepackage{lipsum}
\usepackage{widetext}

\newcommand{\myvec}[1]{\boldsymbol{#1}}
\newcommand{\Mpch}{\,h^{-1}{\rm Mpc}}
\newcommand{\hMpc}{\,h{\rm Mpc}^{-1}}
\newcommand{\Msh}{\,h^{-1}M_\odot}

\newcommand{\rd}{{\rm d}}

\title[Galaxies in braneworld models I]{Galaxy formation in the brane world I: overview and first results}

\author[C.~Hern\'andez-Aguayo et al.]
{
C\'esar Hern\'andez-Aguayo$^{1,2,3}$\thanks{E-mail: 
cesarhdz@mpa-garching.mpg.de (CH-A)},
Christian Arnold$^{1}$\thanks{E-mail: 
christian.arnold@durham.ac.uk (CA)}, 
Baojiu Li$^{1}$ 
and Carlton M. Baugh$^{1}$
\\
\\
$^{1}$Institute for Computational Cosmology, Department of Physics, Durham University, South Road, Durham, DH1 3LE, UK.\\
$^{2}$Max-Planck-Institut f\"ur Astrophysik, Karl-Schwarzschild-Str. 1, D-85748, Garching, Germany\\
$^{3}$Excellence Cluster ORIGINS, Boltzmannstrasse 2, D-85748 Garching, Germany\\
}

\date{Accepted XXX. Received YYY; in original form ZZZ}

\pubyear{2020}

\begin{document}
\label{firstpage}
\pagerange{\pageref{firstpage}--\pageref{lastpage}}
\maketitle

\begin{abstract}
We carry out ``full-physics'' hydrodynamical simulations of galaxy formation in the normal-branch Dvali-Gabadadze-Porrati (nDGP) braneworld model using a new modified version of the {\sc Arepo} code and the IllustrisTNG galaxy formation model. We simulate two nDGP models (N5 and N1) which represent, respectively, weak and moderate departures from GR, in boxes of sizes $62\,h^{-1}{\rm Mpc}$ and $25\,h^{-1}{\rm Mpc}$ using $2\times512^3$ dark matter particles and initial gas cells. This allows us to explore, for the first time, the impact of baryonic physics on galactic scales in braneworld models of modified gravity and to make predictions on the stellar content of dark matter haloes and galaxy evolution through cosmic time in these models. We find significant differences between the GR and nDGP models in the power spectra and correlation functions of gas, stars and dark matter of up to $\sim 25$ per cent on large scales. Similar to their impact in the standard cosmological model ($\Lambda$CDM), baryonic effects can have a significant influence over the clustering of the overall matter distribution, with a sign that depends on scale. Studying the degeneracy between modified gravity and galactic feedback in these models, we find that these two physical effects on matter clustering can be cleanly disentangled, allowing for a method to accurately predict the matter power spectrum with baryonic effects included, without having to run hydrodynamical simulations. Depending on the braneworld model, we find differences compared with GR of up to $\sim15$ per cent in galaxy properties such as the stellar-to-halo-mass ratio, galaxy stellar mass function, gas fraction and star formation rate density. The amplitude of the fifth force is reduced by the presence of baryons in the very inner part of haloes, but this reduction quickly becomes negligible above $\sim0.1$ times the halo radius.
\end{abstract}


%%%%%%%%%%%%%%%%%%%%%%%%%%%%%%%%%%%%%%%%%%%%%%%%%%

%%%%%%%%%%%%%%%%% BODY OF PAPER %%%%%%%%%%%%%%%%%%

%---------------------------------------------------------------
\section{Introduction}
\label{sec:intro}
%---------------------------------------------------------------
Cosmological simulations of galaxy formation are an important instrument to understand the origin, evolution, distribution and properties of galaxies in the Universe. Future galaxy surveys such as the Dark Energy Spectroscopic Instrument \citep{DESI:2016zmz}, Euclid \citep{Laureijs:2011gra} and the Vera C. Rubin Observatory \citep[formerly known as the Large Synoptic Survey Telescope;][]{LSST:2009} aim to measure the position of millions of galaxies to map the large-scale structure of the Universe, a key component to unveil the nature of the dark matter and dark energy, and to test the theory of gravity at an unprecedented level of precision. To fully exploit these data it is essential to provide accurate theoretical predictions for as wide a range of cosmological models as possible.

The standard cosmological model ($\Lambda$ cold dark matter; $\Lambda$CDM) based on Einstein's general relativity (GR) has been widely studied using numerical simulations over the past three decades. In the last five years in particular, hydrodynamical simulations of galaxy formation have been able to model the galaxy population in  cosmological volumes,  achieving encouraging  matches to observations and providing a detailed description of the properties and evolution of galaxies over cosmic time \citep[see, e.g.,][]{Vogelsberger:2014kha,Schaye:2014tpa,Feng:2015jza,McCarthy:2016mry,Pillepich:2017jle,Pillepich:2019bmb,Lee:2020L}.
For instance, the IllustrisTNG (TNG) project is one of the most complete suites of cosmological simulations of galaxy formation to date 
\citep[see e.g.,][]{Pillepich:2017fcc, Nelson:2017cxy, Springel:2017tpz, naiman2018, marinacci2018, Pillepich:2019bmb, Nelson:2019jkf}.
The TNG simulations cover cosmological volumes of $50^3{\rm Mpc}^3$ (TNG50), $100^3{\rm Mpc}^3$ (TNG100) and $300^3{\rm Mpc^3}$ (TNG300). The TNG300 is one of the largest ``full-physics" simulations currently available which allows us to study baryonic effects on the clustering of matter on relatively large-scales.

While $\Lambda$CDM offers a simple yet very successful phenomenological description of most observations, the validity of GR on cosmological scales has so far only been confirmed to relatively poor accuracy compared to tests on smaller scales, and the presence of the cosmic acceleration has motivated the study of alternative theories of gravity that modify GR on large scales \citep{Koyama:2015vza,Ishak:2018his}. The past decade has seen an increasing interest in such theories, leading to a large body of literature on their cosmological behaviours and possible observational tests. However, there has been little work on hydrodynamical simulations of non-standard gravity models \citep[e.g.,][]{Arnold:2013nfr,Arnold:2014qha,hammami2015,He:2015mva,Arnold:2016arp,Ellewsen:2018tww,Arnold:2019vpg}. Exploring the effect of modified gravity on galactic scales in a cosmological context hence remains an important open topic that requires more quantitative work. Hydrodynamical cosmological simulations provide the missing link that connect the properties of dark matter haloes with luminous galaxies.

The Dvali-Gabadadze-Porrati (DGP) braneworld model \citep{Dvali:2000hr} is one of the most widely-studied modified gravity models that employs the Vainshtein screening mechanism \citep{Vainshtein:1972sx}. In this model, normal matter is confined to a 4-dimensional brane embedded in a 5-dimensional spacetime, the bulk. The model leads to two branches of cosmological solutions, dubbed as the {\it self-accelerating} branch (sDGP) and the {\it normal} branch (nDGP). In the sDGP branch, gravity leaks from the brane to the bulk leading to an accelerating expansion without the need to invoke a cosmological constant or dark energy component. However, this model suffers from {\it ghost} instabilities (negative kinetic energy) which results in issues with observational data \cite[see e.g.,][]{Song:2006jk,Fang:2008kc}. On the other hand, the nDGP model does not suffer from ghost instabilities, but it is necessary to include a component of dark energy to match the observed late-time accelerated expansion of the Universe. The nDGP model nevertheless offers the possibility to test the Vainshtein screening mechanism using astrophysical and cosmological probes.

The first numerical simulations of the DGP model were performed by \citet{Schmidt:2009sg,Schmidt:2009sv}, followed by simulations for both the self-accelerated and the normal branches of the DGP model carried out with the adaptive-mesh-refinement (AMR) code {\sc ecosmog-V} \citep{Li:2013nua}. The performance of both codes was tested by \citet{Winther:2015wla}, who found excellent agreement for the prediction of the dark-matter distribution and halo statistics over cosmic time.

To date, the nDGP model has been widely tested using a range of astrophysical and cosmological probes. For example, \citet{Falck:2014jwa,Falck:2015rsa} studied the morphology and the local environmental density of dark matter haloes in the nDGP model. Moreover, \citet{Falck:2017rvl} investigated the effect of the Vainshtein screening mechanism on cosmic voids.  Using a halo occupation distribution (HOD) model,  \citet{Barreira:2016ovx} and \citet{Hernandez-Aguayo:2018oxg} studied galaxy clustering through redshift-space distortions for two different nDGP models. An additional study of cosmic voids in nDGP models was carried out by \citet{Paillas:2018wxs}. More recently, \citet{Devi:2019swk} investigated the galaxy-halo connection and the environmental dependence of the galaxy luminosity function using a subhalo abundance matching approach in modified gravity (including the nDGP model). All the studies mentioned above are dark matter only N-body simulations carried out using the {\sc ecosmog-V} code. Hence, the realisation of full-physics hydrodynamical simulations for galaxy formation is a natural step to continue testing the nDGP model.

Here, we present an extension of the SHYBONE (Simulating HYdrodynamics BeyONd Einstein) simulations \citep{Arnold:2019vpg} by exploring galaxy formation in the nDGP model with an identical expansion history to $\Lambda$CDM \citep{Schmidt:2009sv}. To carry out these simulations, we extended the {\sc Arepo} code \citep{Springel:2009aa} to include the nDGP model and employed its AMR modified gravity solver together with the IllustrisTNG (TNG) galaxy formation model \citep{Weinberger:2017MNRAS,Pillepich:2017jle}. Our simulations represent a further step in the understanding of modified gravity theories on galactic scales.

The first series of the SHYBONE simulations were devoted to studying the interplay between baryonic physics and modifications of gravity in the $f(R)$ gravity model of \citet{Hu:2007nk}. \citet{Arnold:2019vpg} presented the first results on galaxy properties in these models. \citet{Arnold:2019zup} analysed the statistics of matter, haloes and galaxies, making predictions for the matter and halo correlation functions, the halo and galaxy host halo mass functions, the subhalo and satellite galaxy counts, and the correlation function of stars. Using these simulations \citet{Leo:2019ada} studied the effects of modified gravity on the abundance of HI-selected galaxies and their power spectra. The current paper is the first in a series in which we  will present parallel analyses of the SHYBONE-nDGP simulations.

This paper is structured as follows. In Section \ref{sec:nDGP} we introduce the theoretical aspects of the nDGP model. Section \ref{sec:methods} presents our simulation methodology and discusses technical aspects of the numerical implementation. In Section \ref{sec:code_test}, we show some tests to ensure that our implementation works accurately. We describe our SHYBONE-nDGP simulations in Section \ref{sec:S_nDGP}. The first analysis of the new full-physics simulations is presented in Section \ref{sec:Results}. Finally, we summarise the results and give our conclusions in Section \ref{sec:conc}.

%------------Table----------------
\begin{table*}
\caption{Summary of the SHYBONE-nDGP simulations.}
\begin{tabular}{lcccccc}
\hline
Simulation        & Gravity model     & $L_{\rm box}$ $[\Mpch]$ & $N_{\rm DM}$ & $N_{\rm gas}$ & $m_{\rm DM}$ $[\Msh]$ & $m_{\rm gas}$ $[\Msh]$ \\ \hline
L62 & GR, N5, N1 & $62$                    & $512^3$      & $512^3$       & $1.28\times 10^8$      & $2.40\times 10^7$       \\
L25 & GR, N5, N1 & $25$                    & $512^3$      & $512^3$       & $8.41\times 10^6$      & $1.57\times 10^6$       \\
L62-DMO      & GR, N5, N1 & $62$                    & $512^3$      & $512^3$       & $1.52\times 10^8$      & ----                  \\ \hline
\end{tabular}\label{tab:sims}
\end{table*}
%---------------------------------------------------------------
\section{The \lowercase{n}DGP model}
\label{sec:nDGP}
%---------------------------------------------------------------
In the DGP model, the Universe is a 4-dimensional brane embedded in a 5-dimensional bulk spacetime. The gravitational action of the model is given by,
\begin{equation}\label{eq:S_dgp}
S = \int_{\rm brane} {\rm d}^4x \sqrt{-g} \left(\frac{R}{16\pi G}\right) + \int {\rm d}^5x \sqrt{-g^{(5)}} \left(\frac{R^{(5)}}{16\pi G^{(5)}}\right),
\end{equation}
where $g$, $R$ and $G$ correspond to the determinant of the metric, the Ricci scalar and the gravitational constant in the 4-D brane, while $g^{(5)}$, $R^{(5)}$ and $G^{(5)}$ are their respective equivalents in the 5-D bulk. A new parameter, defined by the ratio of $G^{(5)}$ and $G$, is known as the crossover scale, $r_{\rm c}$,
\begin{equation}\label{eq:rc}
r_{\rm c} = \frac{1}{2} \frac{G^{(5)}}{G}\,.
\end{equation}
Here we study the normal branch (nDGP) model, where the variation of the action, Eq.~\eqref{eq:S_dgp}, yields  the modified Friedmann equation 
\begin{equation}\label{eq:H_ndgp}
\frac{H(a)}{H_0} = \sqrt{\Omega_{\rm m}a^{-3} + \Omega_{\rm DE}(a) + \Omega_{\rm rc}} - \sqrt{\Omega_{\rm rc}},
\end{equation}
in a homogeneous and isotropic universe with $\Omega_{\rm rc} = c^2/(4H^{2}_{0}r^2_{\rm c})$ where $c$ is the speed of light, $\Omega_{\rm m}$ is the present-day value of the matter density parameter, the dark energy density parameter $\Omega_{\rm DE}(a)$ is defined as $\Omega_{\rm DE}(a) \equiv 8\pi G\rho_{\rm DE}(a)/3H^2(a)$, $a$ is the scale factor and $H_0$ is the present-day value of the Hubble parameter. In this model, deviations from GR can be characterised in terms of the parameter $H_0 r_{\rm c} / c$. As we can see from Eq.~\eqref{eq:H_ndgp} if $H_0 r_{\rm c} / c \rightarrow \infty$ then the expansion of the universe is closer to $\Lambda$CDM. Therefore, we consider two nDGP models with $H_0r_c / c = 5$, hereafter referred to as N5  and $H_0 r_{\rm c} / c = 1$ as N1; N5 represents a weak departure from GR, whereas N1 corresponds to a medium departure.

%---------------------------------------------------------------
\subsection{Structure formation in the nDGP model}
\label{sec:structure_nDGP}
%---------------------------------------------------------------
Structure formation in the nDGP model is governed by the following equations in the quasi-static and weak-field limits \citep{Koyama:2007ih},
\begin{equation}\label{eq:poisson_nDGP}
\nabla^2 \Phi = 4\pi G a^2 \delta \rho_{\rm m} + \frac{1}{2}\nabla^2\varphi\,,
\end{equation}
\begin{equation}\label{eq:phi_dgp}
\nabla^2 \varphi + \frac{r_{\rm c}^2}{3\beta\,a^2c^2} \left[ (\nabla^2\varphi)^2
- (\nabla_i\nabla_j\varphi)^2 \right] = \frac{8\pi\,G\,a^2}{3\beta} \delta\rho_{\rm m}\,,
\end{equation}
where $\nabla^2$ is the three-dimensional Laplacian operator, $\varphi$ is a scalar degree of freedom related to the bending mode of the brane, $\delta\rho_{\rm m} = \rho_{\rm m} - \bar{\rho}_{\rm m}$ is the perturbation of non-relativistic matter density, and
\begin{equation}\label{eq:beta_dgp}
\beta = 1 + 2 H\, r_{\rm c} \left ( 1 + \frac{\dot H}{3 H^2} \right )
= 1 + \frac{\Omega_{\rm m}a^{-3} + 2\Omega_\Lambda}{2\sqrt{\Omega_{\rm rc}(\Omega_{\rm m}a^{-3} + \Omega_{\Lambda})}}.
\end{equation}
In the last expression we have assumed a $\Lambda$CDM background.

If we linearise Eq.~\eqref{eq:phi_dgp}, the two nonlinear terms in the squared brackets vanish and the modified Poisson equation, Eq.~\eqref{eq:poisson_nDGP}, can be re-expressed as 
\begin{equation}\label{eq:dgp_linear}
 \nabla^2 \Phi = 4\pi G a^2 \left( 1 + \frac{1}{3\beta} \right) \delta\rho_{\rm m},
\end{equation}
which represents a time-dependent and scale-independent rescaling of Newton's constant. Since $\beta$ is always positive, the formation of structure is enhanced in this model with respect to $\Lambda$CDM.
%---------------------------------------------------------------
\subsection{Vainshtein mechanism}
\label{sec:screen}
%---------------------------------------------------------------
The nDGP model is a representative class of modified gravity models that feature the Vainshtein screening mechanism \citep{Vainshtein:1972sx}. To illustrate how the Vainshtein mechanism works, let us for simplicity consider solutions in spherical symmetry, where Eq.~\eqref{eq:phi_dgp} can be written in the following form
\begin{equation}\label{eq:dgp}
\frac{2r_{\rm c}^2}{3\beta c^2 a^2}\frac{1}{r^2}\frac{\rd}{\rd r}\left[r\left(\frac{\rd\varphi}{\rd r}\right)^2\right] + \frac{1}{r^2}\frac{\rd}{\rd r}\left[r^2\frac{\rd\varphi}{\rd r}\right]
= \frac{8\pi G}{3\beta}\delta\rho_{\rm m}a^2\,.
\end{equation}
Defining the mass enclosed in radius $r$ as
\begin{equation}\label{eq:Mass_r}
M(r) \equiv 4\pi\int^r_0\delta\rho_{\rm m}(r')r'^2{\rd}r',
\end{equation}
we can rewrite Eq.~(\ref{eq:dgp}) as
\begin{equation}\label{eq:dgp_eqn}
\frac{2r_{\rm c}^2}{3\beta c^2}\frac{1}{r}\left(\frac{\rd\varphi}{\rd r}\right)^2 + \frac{\rd\varphi}{\rd r} = \frac{2}{3\beta}\frac{GM(r)}{r^2}\ \equiv\ \frac{2}{3\beta}g_{\rm N}(r),
\end{equation}
in which for simplicity we have set $a=1$, and $g_{\rm N}$ is the Newtonian acceleration caused by the mass $M(r)$ at distance $r$ from the centre, Eq.~\eqref{eq:Mass_r}. 

If we assume that $\delta\rho_{\rm m}$ is a constant within radius $R$ and zero outside,  then Eq.~\eqref{eq:dgp_eqn} has the physical solution
\begin{equation}\label{eq:dvarphidr_in}
\frac{\rd\varphi}{\rd r} = \frac{4}{3\beta}\frac{r^3}{r_{\rm V}^3}\left[\sqrt{1+\frac{r_{\rm V}^3}{r^3}}-1\right]g_{\rm N}(r),
\end{equation}
for $r\geq R$ and
\begin{equation}\label{eq:dvarphidr_out}
\frac{\rd\varphi}{\rd r} = \frac{4}{3\beta}\frac{R^3}{r_V^3}\left[\sqrt{1+\frac{r_V^3}{R^3}}-1\right]g_{\rm N}(r),
\end{equation}
for $r\leq R$. In these expressions $r_{\rm V}$ is the Vainshtein radius which can be written as
\begin{equation}\label{eq:r_V}
r_{\rm V} \equiv \left(\frac{8r^2_{\rm c} r_{\rm S}}{9\beta^2}\right)^{1/3} = \left(\frac{4GM(R)}{9\beta^2H^2_0\Omega_{\rm rc}}\right)^{1/3}\,,
\end{equation}
where $r_{\rm S} \equiv2GM(R)/c^2$ is the Schwarzschild radius and $M(R) \equiv 4\pi\int^R_0\delta\rho_{\rm m}(r')r'^2{\rd}r'$.

According to Eq.~\eqref{eq:poisson_nDGP}, the fifth force is given by $\frac{1}{2}\rd\varphi/\rd r$. Thus when $r\gg r_{\rm V}$ we have
\begin{equation}\label{eq:fifth-force-ratio-asymptotic}
\frac{1}{2}\frac{\rd\varphi}{\rd r} \rightarrow \frac{1}{3\beta}g_N(r),
\end{equation}
meaning on scales larger than the Vainshtein radius gravity is enhanced (because $\beta>0$ for the normal branch of the DGP model). On the other hand, for $r,R\ll r_{\rm V}$ we have
\begin{equation}
\frac{1}{2}\frac{\rd\varphi}{\rd r} \rightarrow \frac{2}{3\beta}\frac{R^{3/2}}{r^{3/2}_{\rm V}}g_{\rm N}(r) \ll g_{\rm N}(r),
\end{equation}
indicating that the fifth force is suppressed (or screened) well within the Vainshtein radius.

\begin{figure*}
\centering
\includegraphics[width=0.32\linewidth]{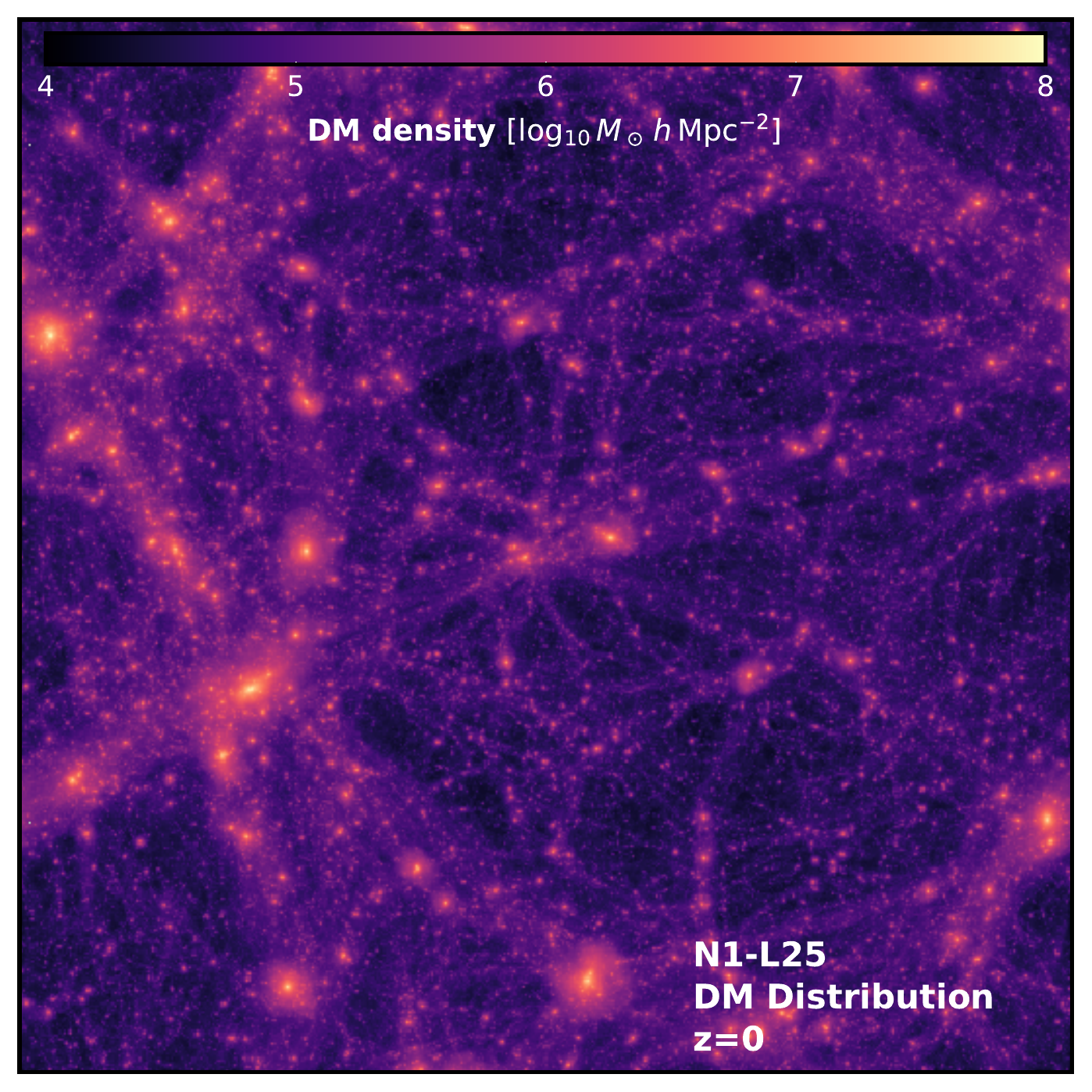}
\includegraphics[width=0.32\linewidth]{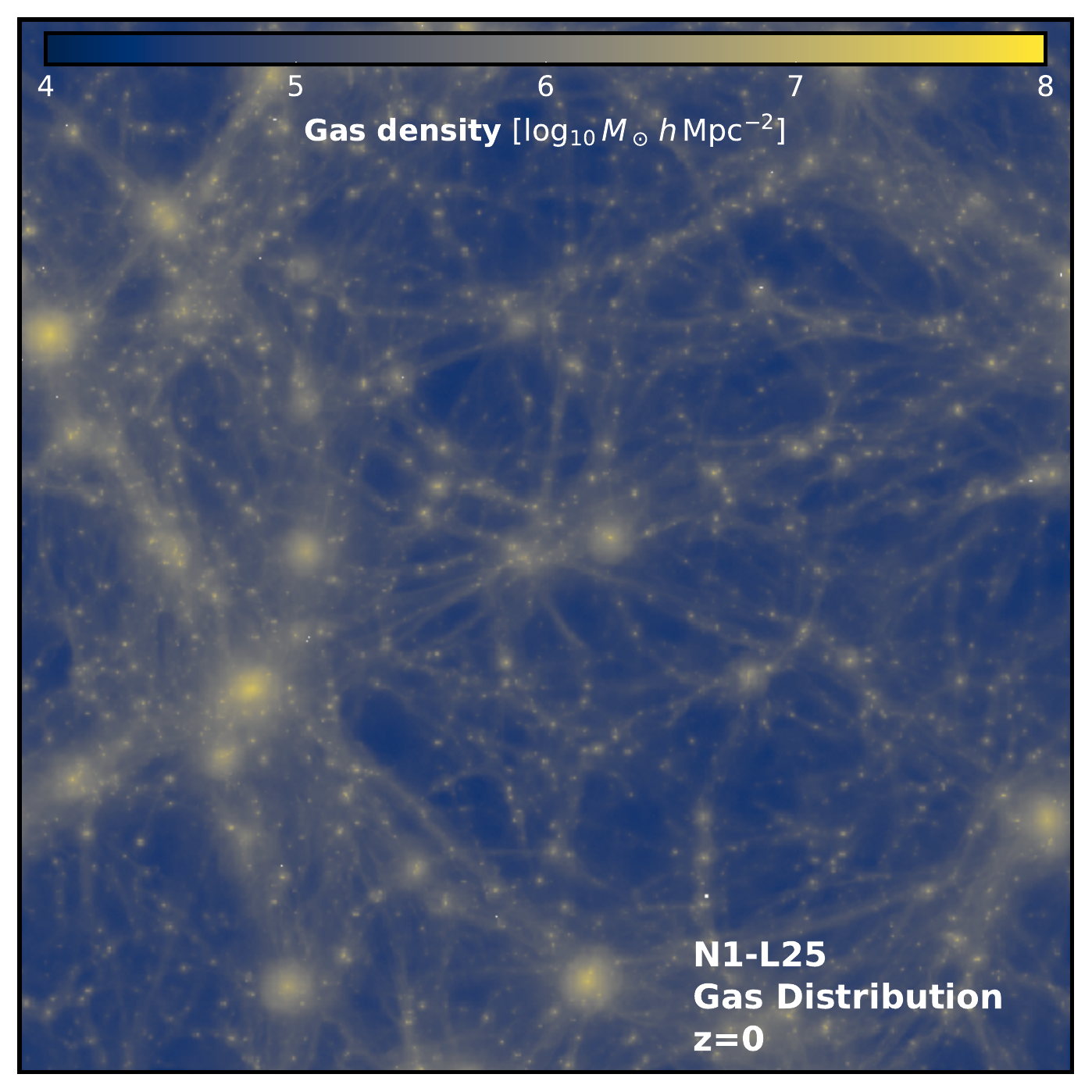}
\includegraphics[width=0.32\linewidth]{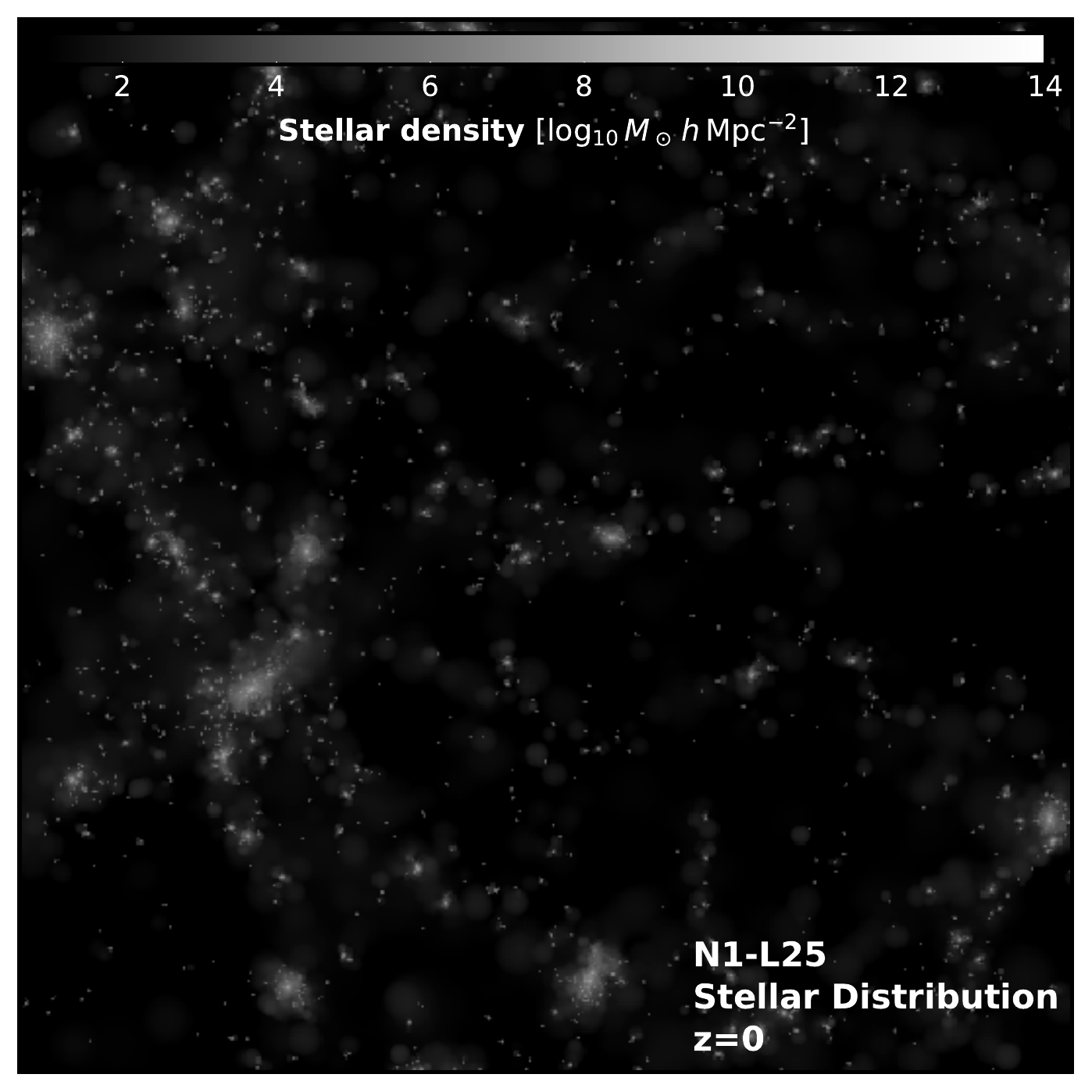}
\includegraphics[width=0.32\linewidth]{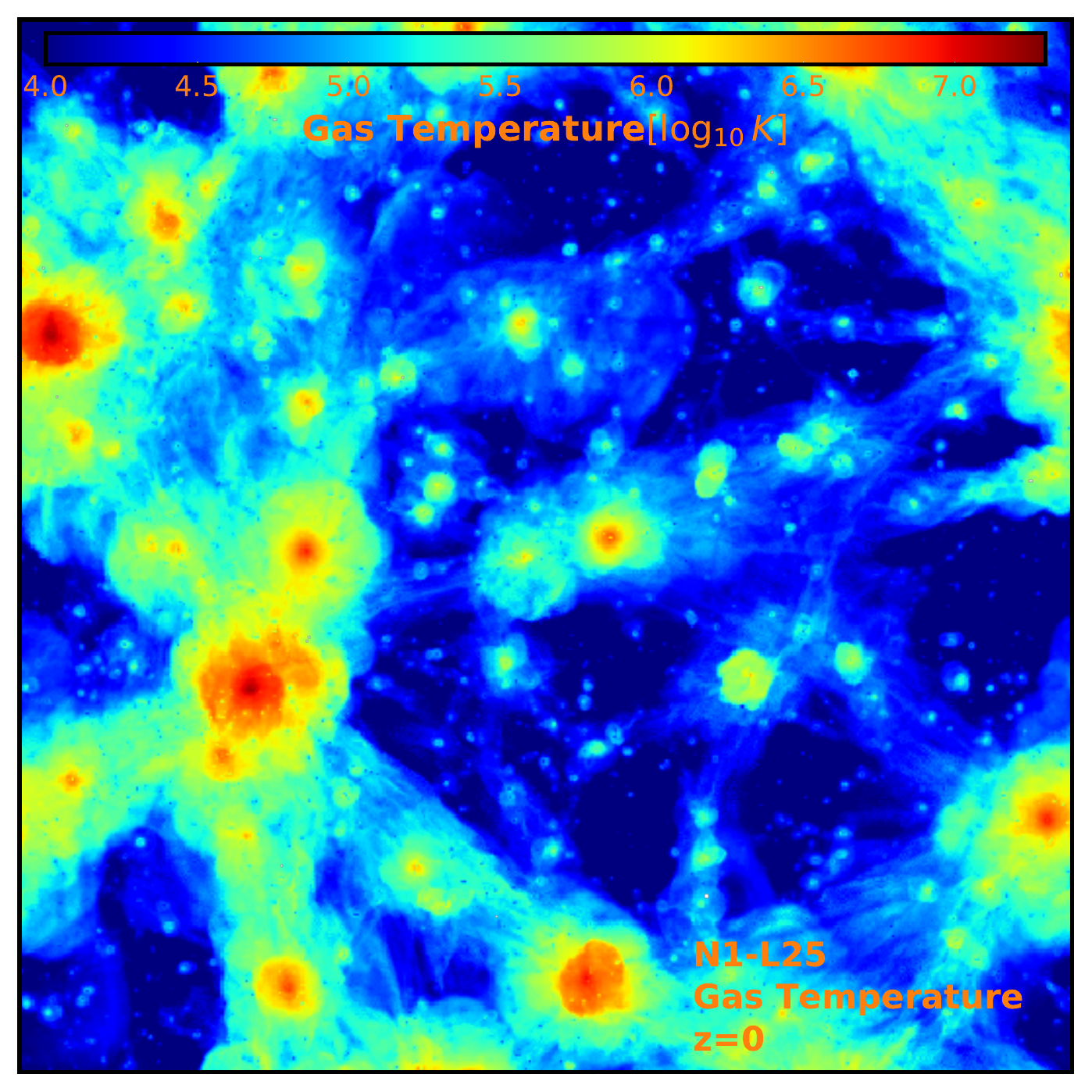}
\includegraphics[width=0.32\linewidth]{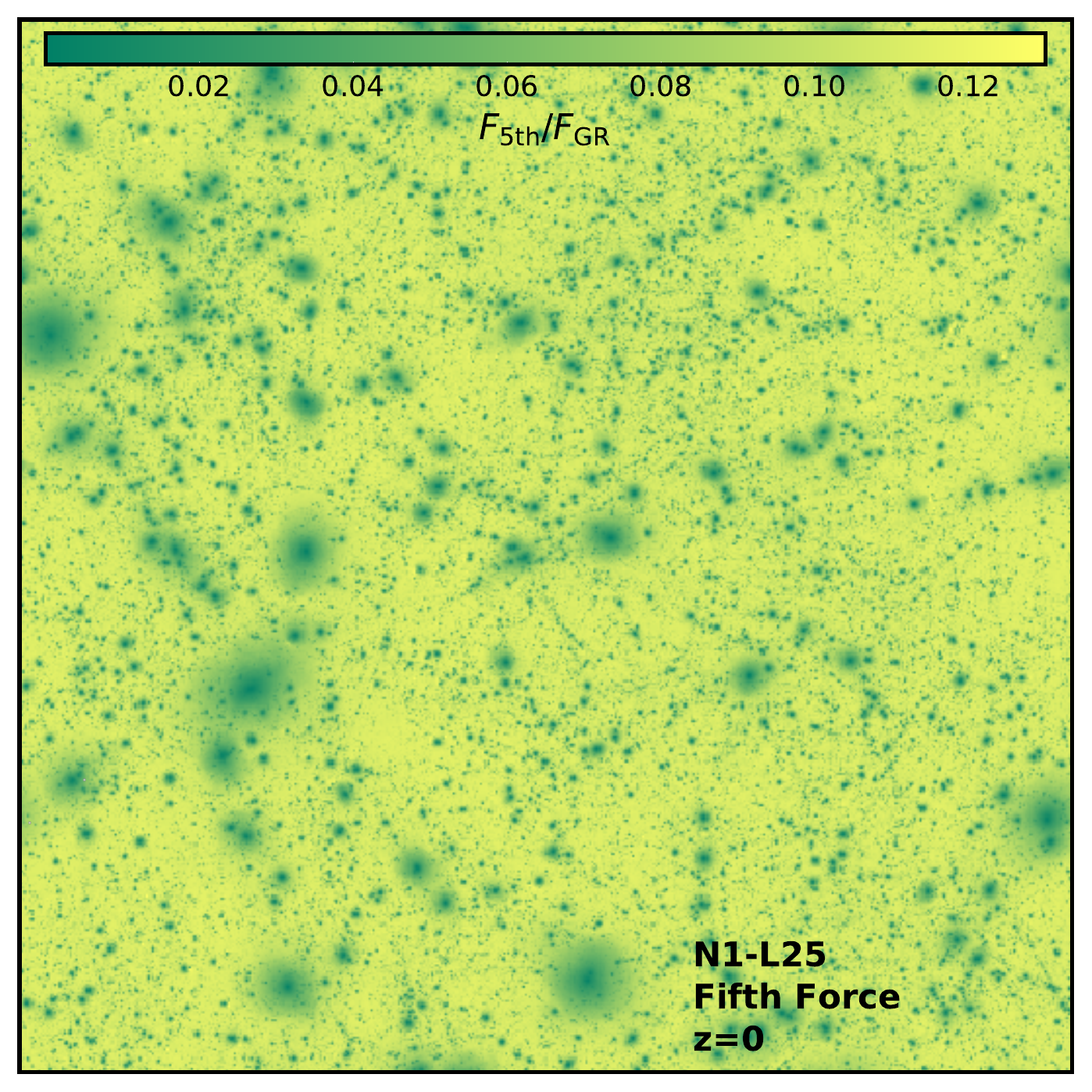}
\includegraphics[width=0.32\linewidth]{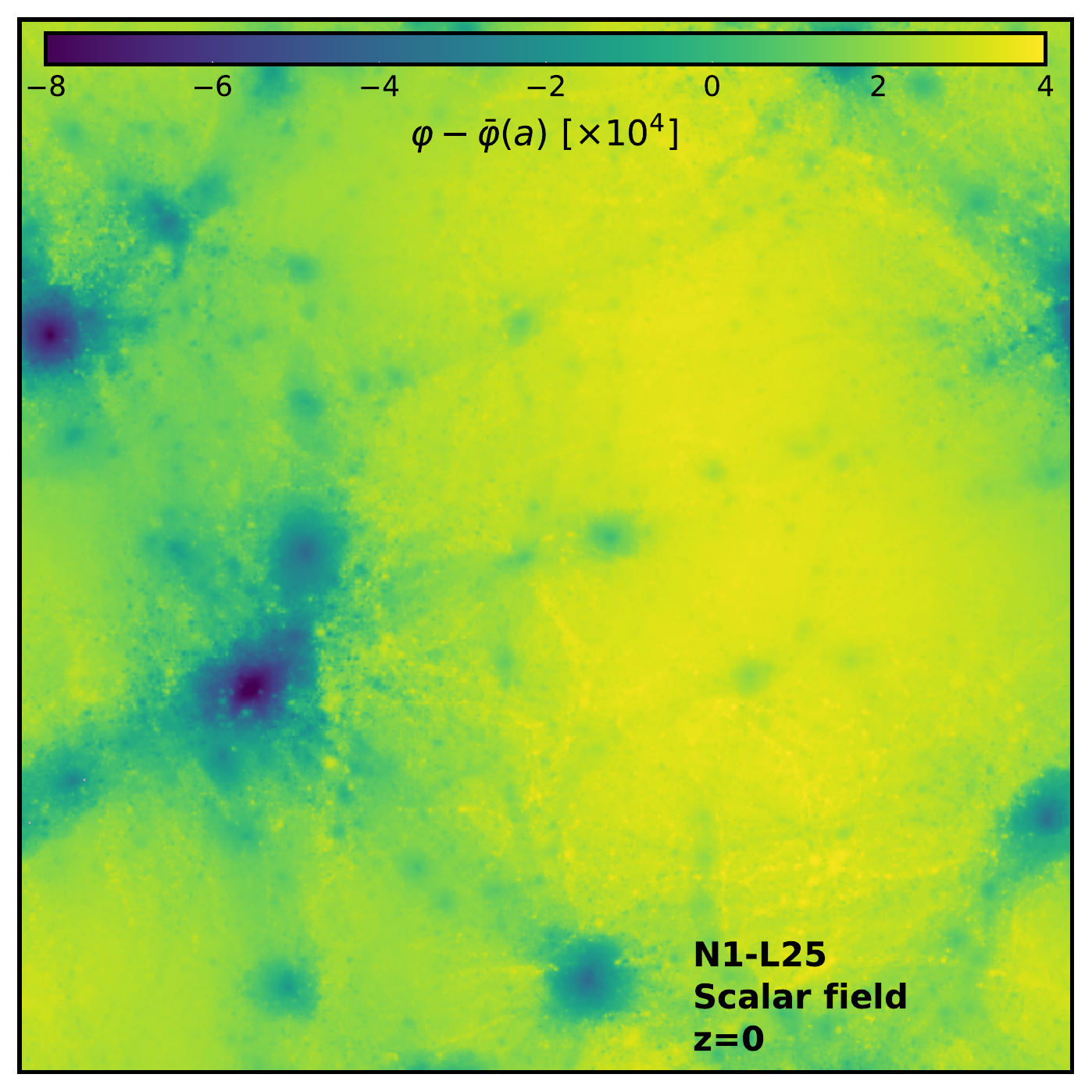}
\includegraphics[width=0.32\linewidth]{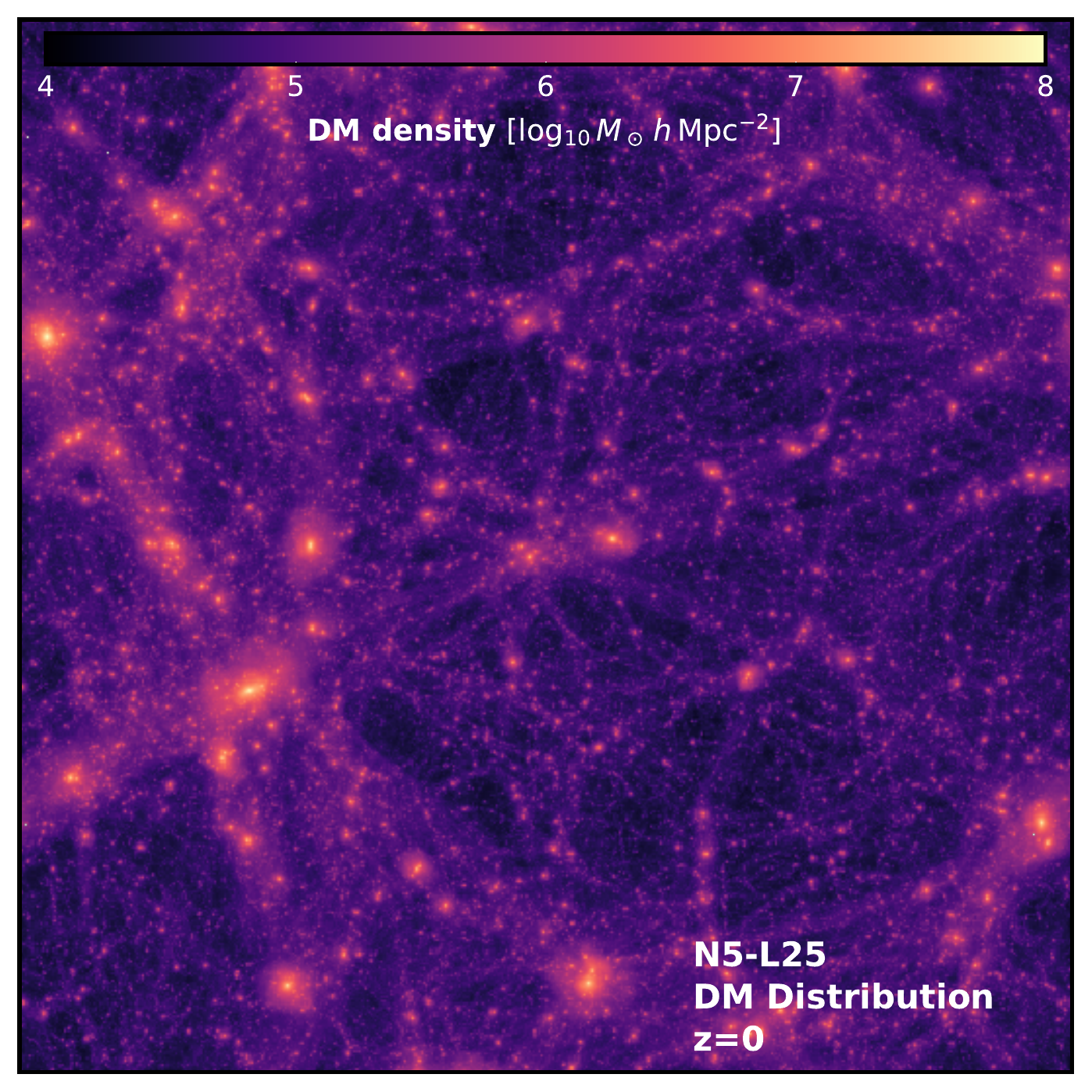}
\includegraphics[width=0.32\linewidth]{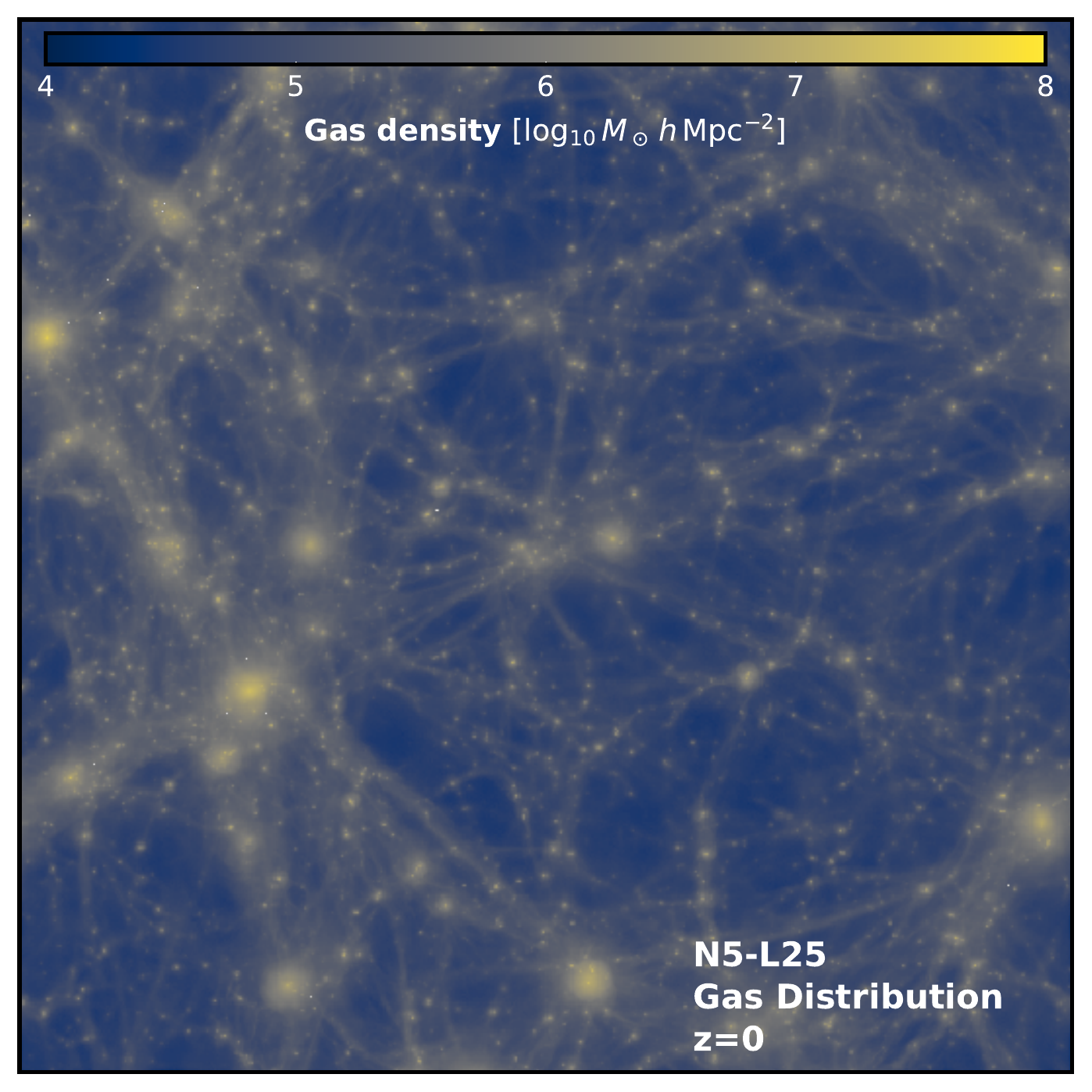}
\includegraphics[width=0.32\linewidth]{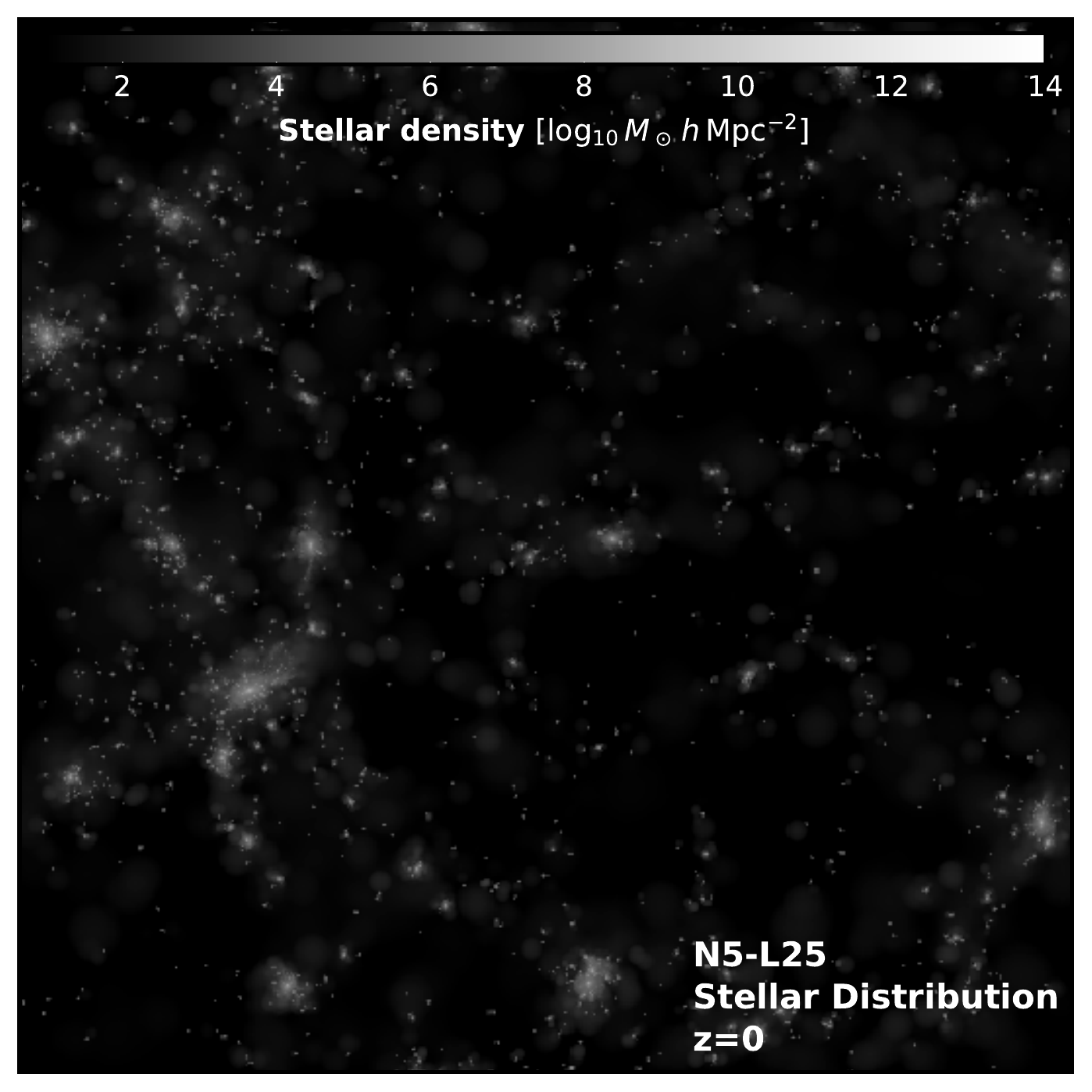}
\includegraphics[width=0.32\linewidth]{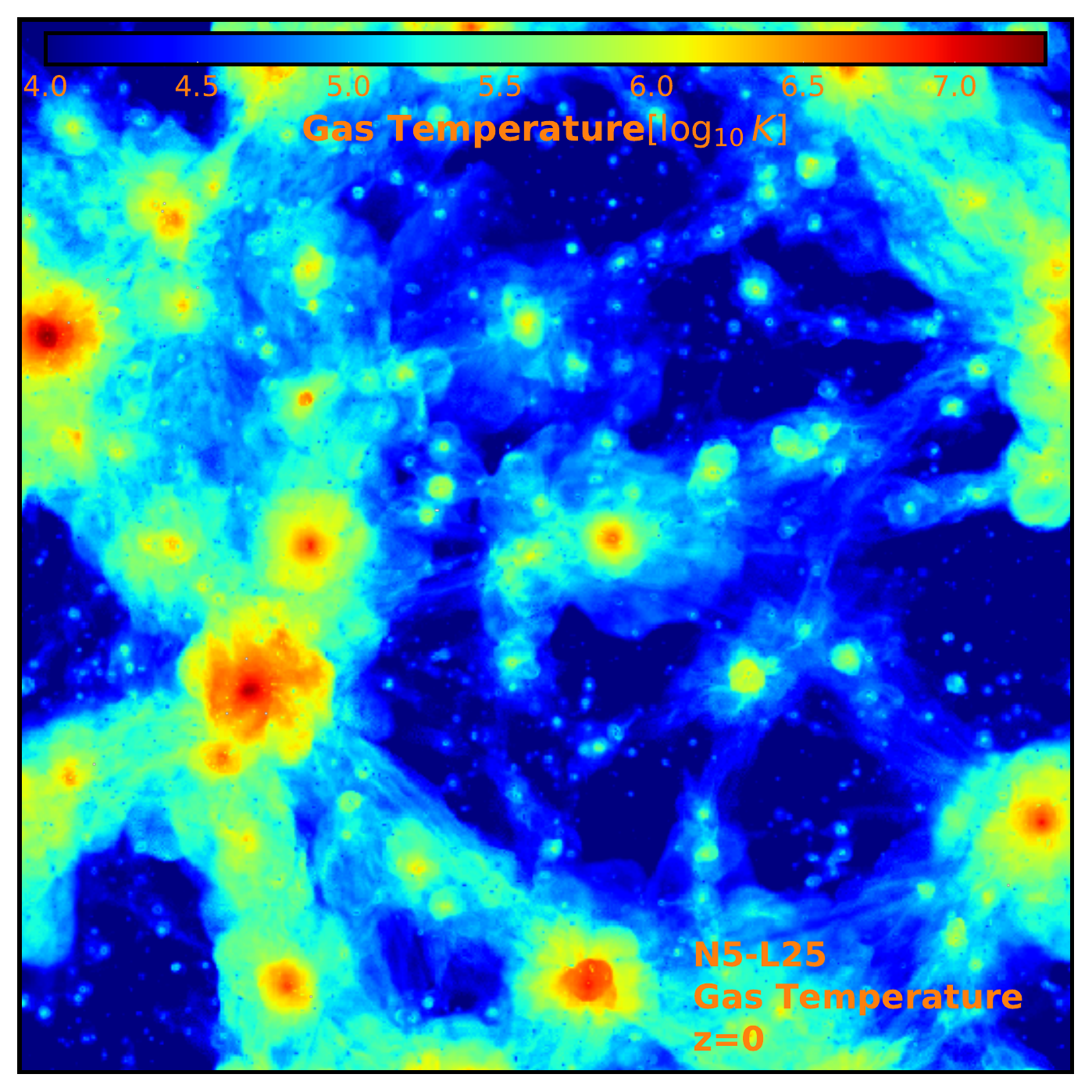}
\includegraphics[width=0.32\linewidth]{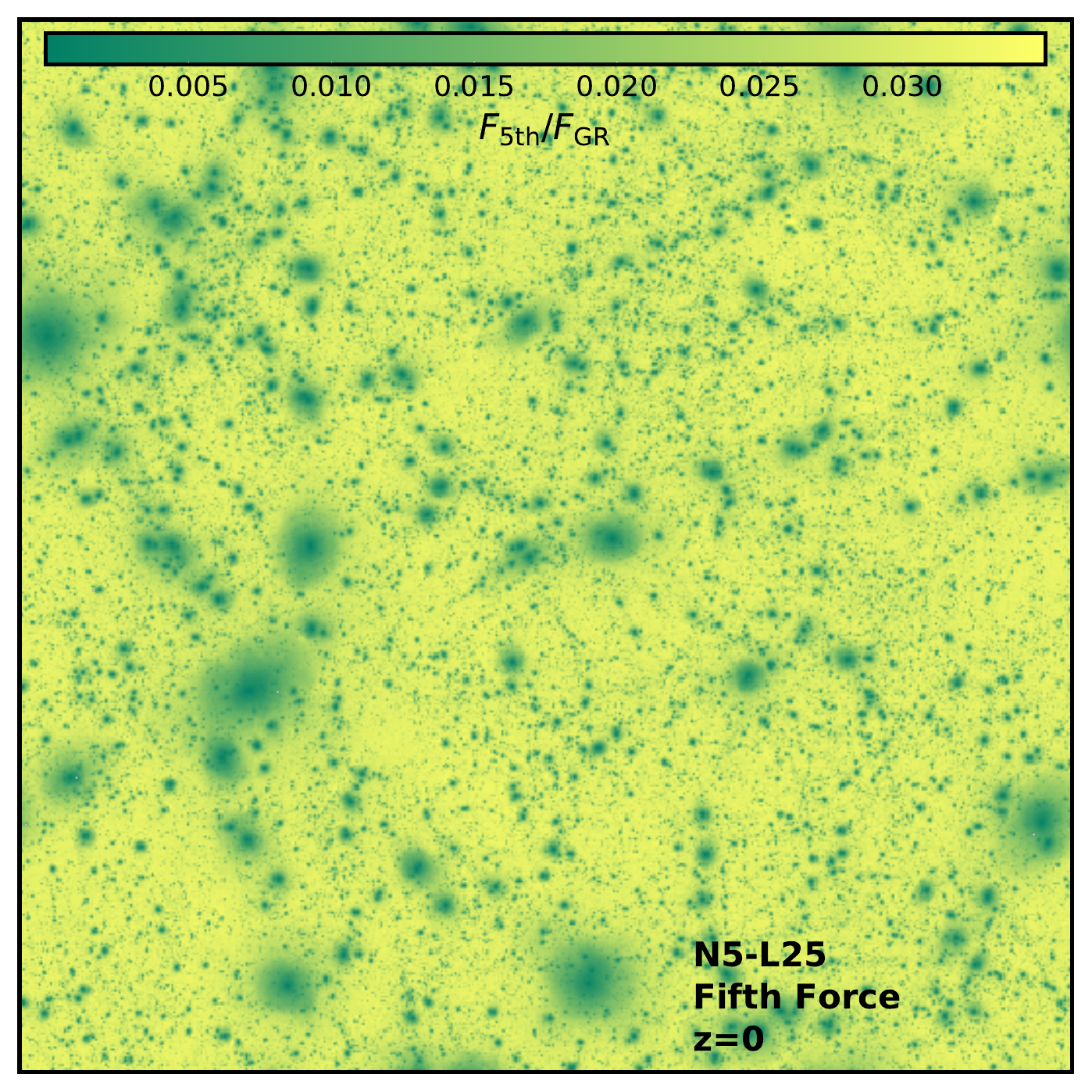}
\includegraphics[width=0.32\linewidth]{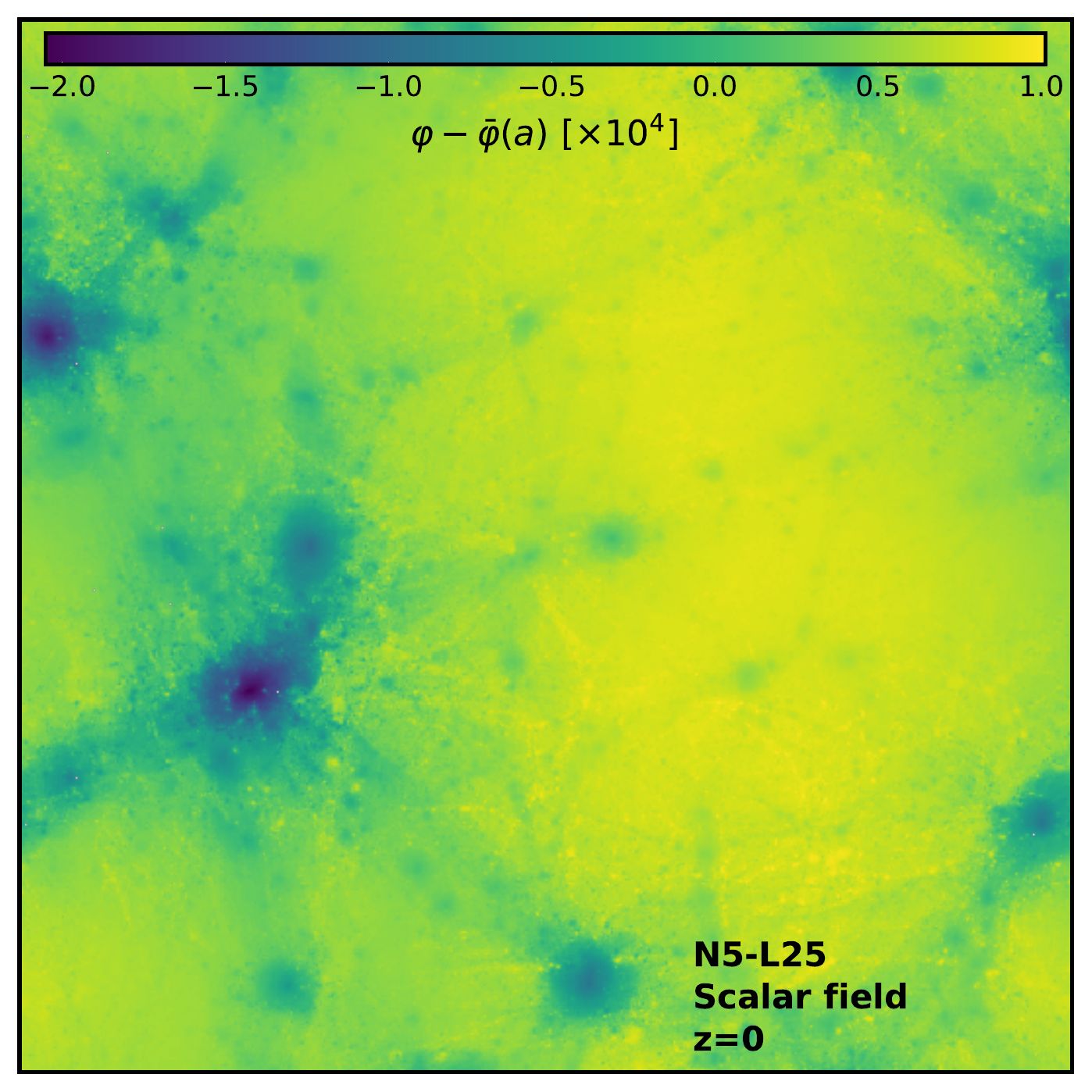}
\caption{Visual inspection of the nDGP-L25 simulations showing the large-scale structures at $z=0$. The {\it top six panels} show the column density of DM, the gas density, the stellar mass distribution, the temperature of the gas, the fifth to Newtonian force ratio, and the difference between the local and background mean values of the scalar field, $\varphi-\bar{\varphi}$ (in code unit), of the N1 model. The {\it bottom six panels} display the same matter and modified gravity quantities but for the N5 model.}
\label{fig:visual}
\end{figure*}

%---------------------------------------------------------------
\section{The SHYBONE \lowercase{n}DGP simulations}\label{sec:S_nDGP}
%---------------------------------------------------------------
The SHYBONE-nDGP runs consist of a suite of nine simulations covering three gravity models (GR, N5 and N1) at two resolutions. The larger box has a size of $L=62\Mpch$ (L62) and contains $512^3$ dark-matter particles and $512^3$ gas elements, giving a mass resolution of $m_{\rm DM} = 1.28\times10^8\Msh$ and $m_{\rm gas} = 2.40\times10^7\Msh$. We have also run a smaller box with size $L=25\Mpch$ (L25) and $2\times512^3$ resolution elements giving a baryon mass resolution of $1.57\times10^6\Msh$ and dark matter particle mass of $8.41\times10^6\Msh$. In addition, we ran DM-only versions of the L62 runs (L62-DMO); in this case the mass of the dark matter particle is $1.52\times10^8\Msh$. The softening lengths for DM particles and stars are $1.25$ and $0.5\,h^{-1}{\rm kpc}$ for the L62 and L25 runs, respectively. Table~\ref{tab:sims} summarises the set-up of our simulations.

For all simulations, we use the same linear perturbation theory power spectrum to generate the  initial conditions at $z_{\rm ini}=127$ with the {\sc N-GenIC} code \citep{Springel:2005nw}, which allies the Zeldovich approximation. The cosmological parameters are chosen from those reported by \citet{Ade:2015xua}: 
$$\{\Omega_{\rm b}, \Omega_{\rm m}, h, n_s, \sigma_8\} = \{0.0486,0.3089,0.6774,0.9667,0.8159\}.$$
For each set of GR, N5 and N1 simulations we use the same initial condition, since at $z=127$ the effect of modified gravity is expected to be negligible, and the initial power spectrum depends only on the other cosmological parameters.

The modified gravity solver is combined with the TNG galaxy formation model \citep{Weinberger:2017MNRAS,Pillepich:2017jle} to follow the formation and evolution of realistic synthetic galaxies through cosmic time. The TNG model uses the Eddington ratio as the criterion for deciding the accretion state of black holes (BHs). In addition, it employs a kinetic AGN feedback model which produces a BH-driven wind --- this is responsible for the quenching of star formation in galaxies residing in intermediate and high-mass dark matter haloes, and for the production of a population of red and passive galaxies at late times \citep{Weinberger:2017MNRAS}. We have not retuned the TNG subgrid physics models for use in the nDGP simulations, and instead use the same galaxy formation prescription for all gravity models. In principle, such a retuning is needed for any new model of gravity, but as we shall see below, the default TNG model still predicts baryonic observables in good agreement with observations; by not recalibrating the TNG model, we can isolate any changes that are due to the effect of gravity on halo formation and structure. Finally, we do not expect the fifth force to have a strong impact on the behaviour of stars and BHs \citep[see, e.g., Section 9.6 of][for a review.]{Baker:2019gxo}.

The dark matter haloes (groups) and their substructures -- subhaloes and galaxies -- are identified with {\sc Subfind}  \citep{Springel:2000qu}. The group catalogues (including subhalo and galaxy information) and the particle data are stored in 100 snapshots from $z\sim20$ to $z=0$. The large number of snapshots is ideal for generating halo merger trees which allows to run with semi-analytic models of galaxy formation. This will be addressed in a future paper. 

%--------- Figure --------------
\begin{figure*}
 \centering
\includegraphics[width=0.9\textwidth]{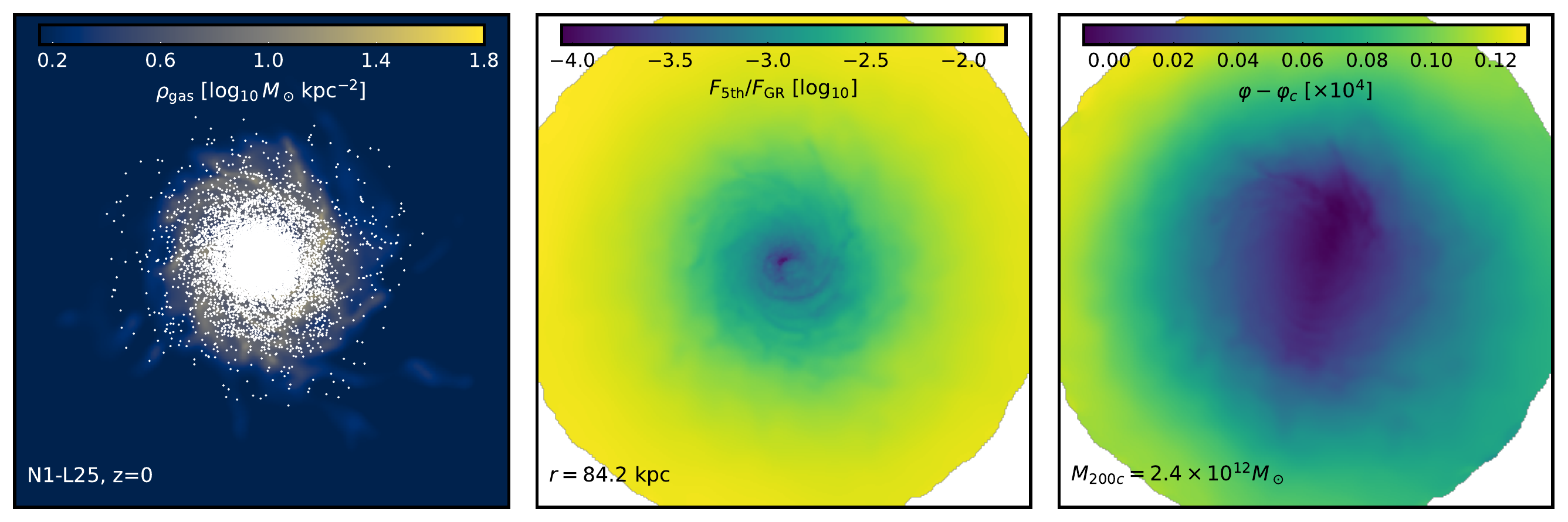}
\includegraphics[width=0.9\textwidth]{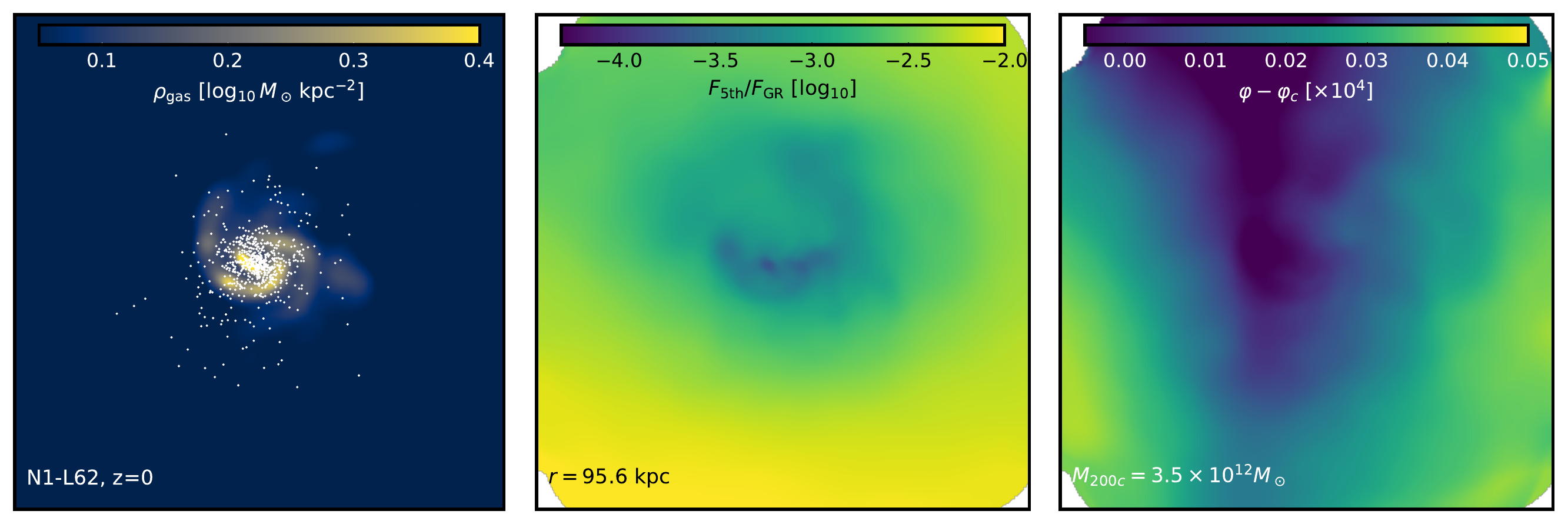}
\includegraphics[width=0.9\textwidth]{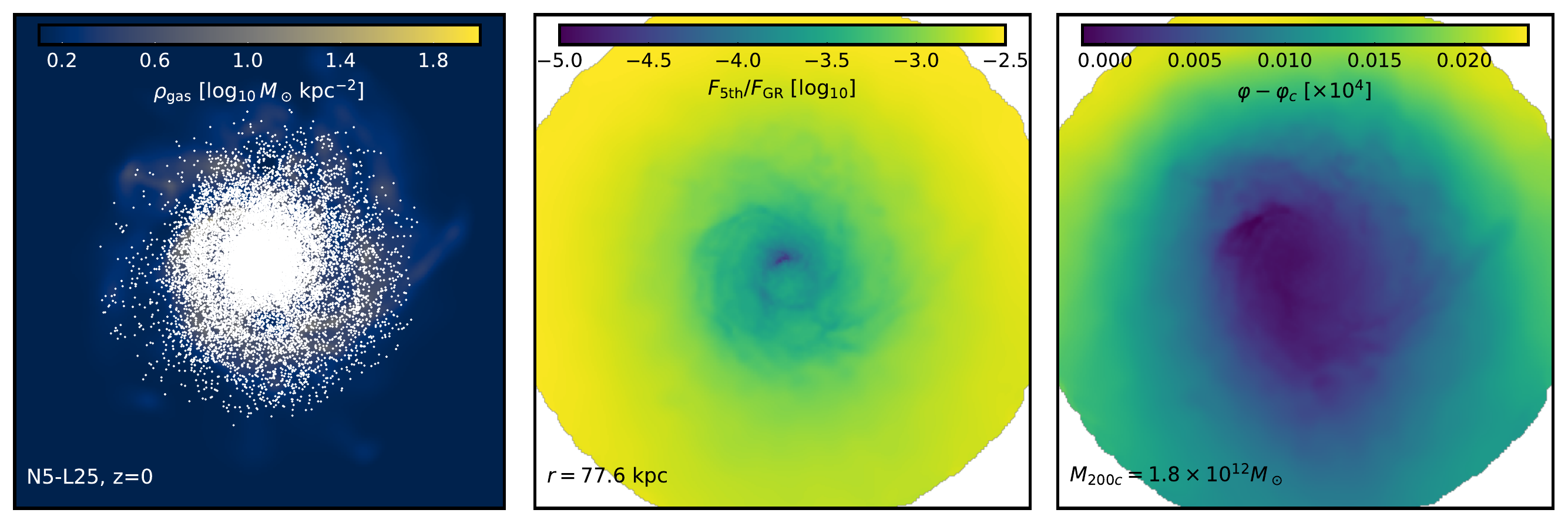}
\includegraphics[width=0.9\textwidth]{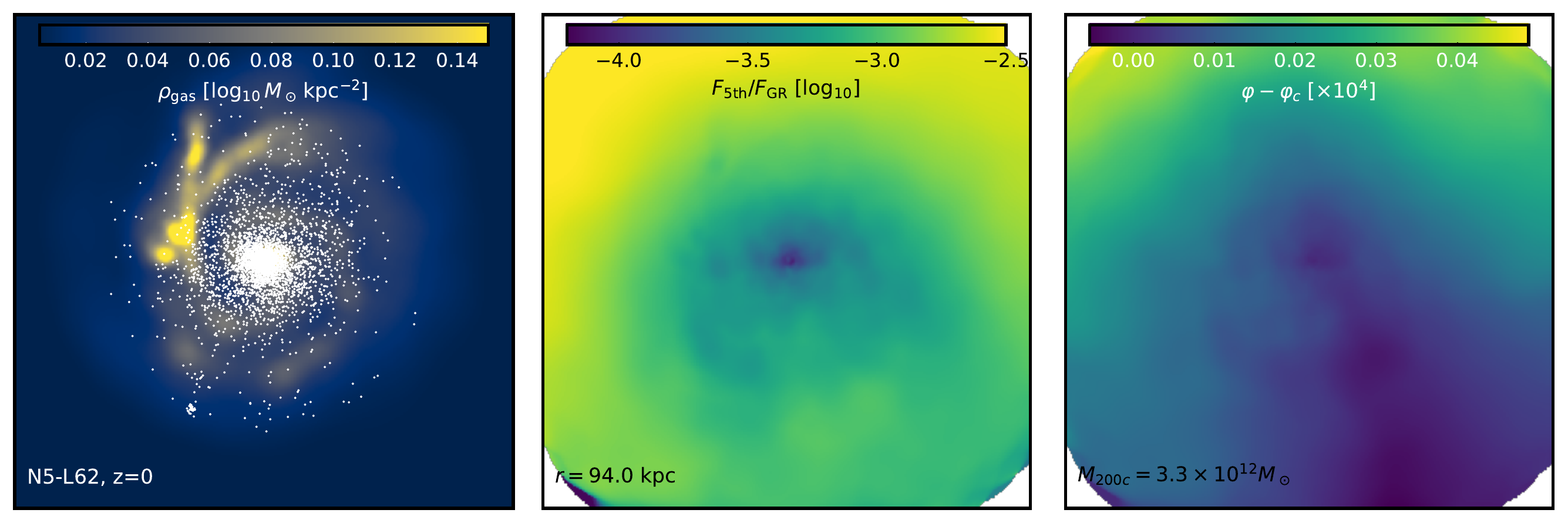}
\caption{Visualisation of a selection of disc galaxies from the two boxes, for both N1 (top two rows) and N5 (bottom two rows). {\it Left column}: the gas column density with stars (white dots) overplotted. {\it Central column}: the ratio between the magnitudes of the fifth force and standard gravity. {\it Right column}: the difference between the local value of the scalar field, $\varphi$, and its value at the galactic centre, ${\varphi}_c$. All galaxies are selected at $z=0$ and all images are face-on. Numerical values are colour-coded as indicated by the colour bars in each panel, and various information, such as the disc radius and host halo mass, is also shown.}
\label{fig:discs}
\end{figure*}

%--------- Figure --------------
\begin{figure*}
 \centering
\includegraphics[width=1\textwidth]{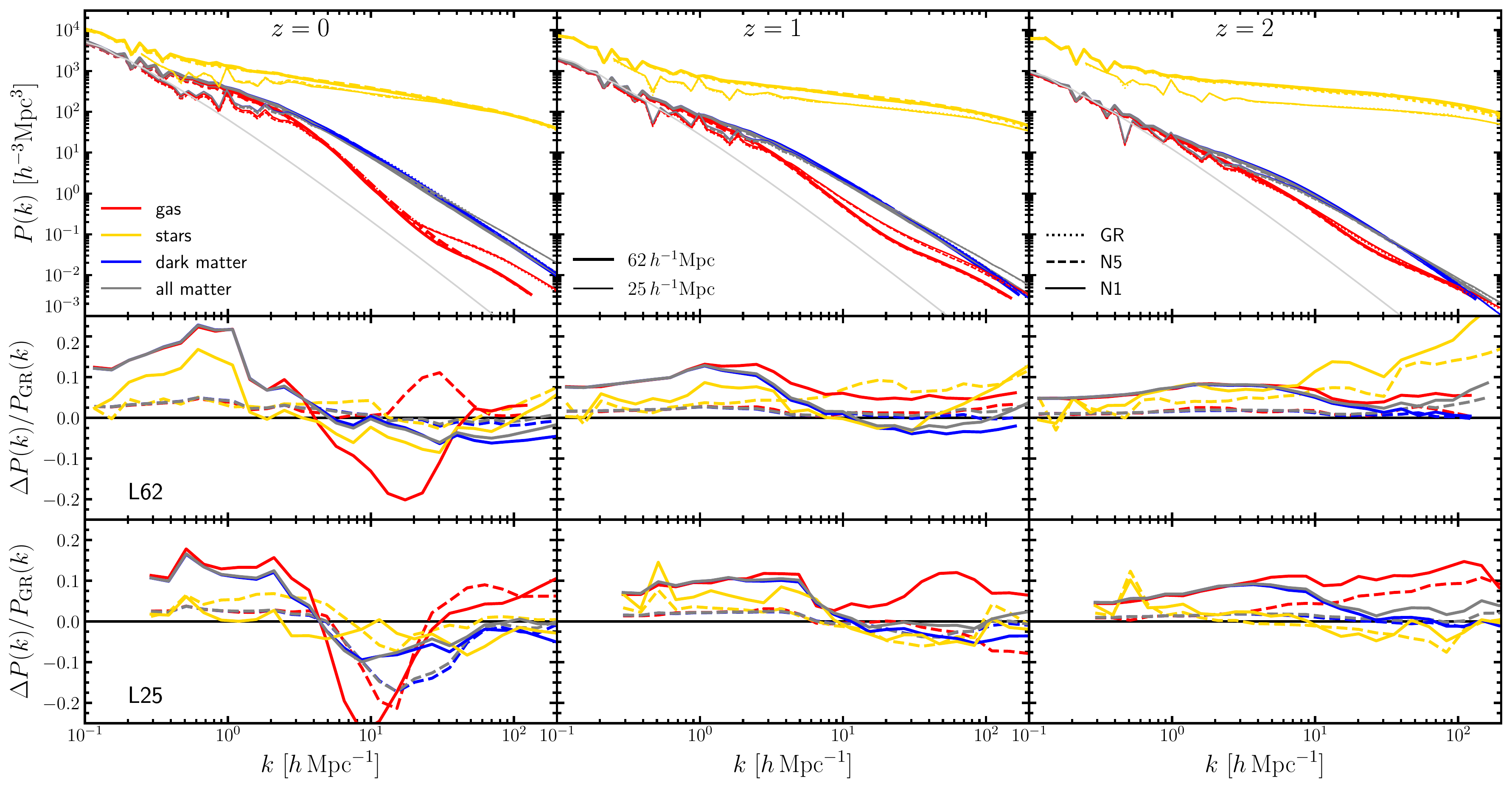}
\caption{The measured power spectra of different matter components of our full-physics simulations at $z=0$ ({\it left panel}), $z=1$ ({\it middle panel}) and $z=2$ ({\it right panel}). The {\it upper panels} show the absolute values of the power spectra of the gas (red lines), stars (yellow lines), dark matter (blue lines) and total matter (grey lines) components. Thick lines indicate the results from the $L=62\Mpch$ box, while thin lines show the results from the $L=25\Mpch$ case. \textit{Dotted lines} show results for GR, \textit{dashed lines} for N5 and \textit{solid lines} for N1. The {\it middle} and {\it lower rows} display the relative differences with respect to GR predictions for the $L=62\Mpch$ and $L=25\Mpch$ boxes, respectively.}
\label{fig:Pk_all}
\end{figure*}

A visual representation of the nDGP-L25 simulations at the present time is displayed in Fig.~\ref{fig:visual}. The top (bottom) six panels show the densities of dark matter, gas and stars, gas temperature, the ratio between the amplitudes of the fifth and standard Newtonian forces, and the scalar field $\varphi$ in the N1 (N5) model. The colour maps were generated with the {\sc SPHviewer} package \citep{ref:sphviewer}. The distribution of matter in our nDGP simulations seems indistinguishable between N1 and N5 (we have not shown the GR results as they are also indistinguishable visually), but we will quantify the impact of modified gravity on the clustering of matter components and on the galaxy properties in the following section. From the force ratio panels of Fig.~\ref{fig:visual} we can observe that high-density (green) regions experience a negligible force enhancement while low-density (yellow) regions experience an enhancement of $F_{\rm 5th} = (1/3\beta)F_{\rm GR}$, where $\beta = 2.69$ and $\beta = 9.45$ [cf.~Eq.~\eqref{eq:beta_dgp}] for N1 and N5 at $z=0$, respectively. In the scalar field panels of Fig.~\ref{fig:visual} we have subtracted the mean scalar field value measured in the whole simulation box, $\bar{\varphi}$. $\varphi-\bar{\varphi}$ then has a zero mean and can be regarded as the potential of the fifth force: as expected, this map is smoother and dominated by long-wavelength modes. Notice that the colour bars for the force ratio and scalar field panels are different between N5 and N1.

Figure~\ref{fig:discs} displays the face-on images of a random\footnote{Note that this means that the galaxies are not matched across the different gravity models, which is difficult to do with a small sample. This is however not a problem as we are not making quantitative comparisons of the galaxies' properties in different models here.} selection of four different disc galaxies from the nDGP full-physics simulations at $z=0$. The first two rows show, respectively, one galaxy from each of the L25 and L62 runs for N1, while the last two rows show two disc galaxies for N5 (again one per box). We follow the prescription of \citet{Arnold:2019vpg} to identify disc galaxies in our simulations. Essentially, we select galaxies that have $\kappa > 0.57$, with $\kappa$ the rotational-to-total kinetic energy parameter \citep{Ferrero:2017M6F}. The first column displays the gas column density (the colour map) and star particles (white dots) of the galaxies. The second column presents the map of modified gravity force enhancement. The amplitude of the scalar field is shown in the third column. We see that the L25 box, owing to its higher resolution, gives rounder and more detailed galaxy images. The fifth force is indeed much weaker than Newtonian gravity inside and around the galactic disc: the ratio between their magnitudes is smaller than $10^{-2}$ and $10^{-2.5}$, for N1 and N5 respectively, in this region, showing that the Vainshtein mechanism effectively suppresses the fifth force. In the scalar field maps, we subtract $\varphi_c$, the scalar field value at the centre of the galaxy (note that this is different from Fig.~\ref{fig:visual}), to eliminate the contribution from long-wavelength modes. This is because we want to see the spatial variations of the scalar field that are caused by the matter distribution in the galaxy itself. We observe that the scalar field increases in value from inside out, as expected given its role as the fifth force potential.

We have counted the number of disc galaxies from the simulations. The L25 runs produce $124$, $126$ and $118$ such objects at $z=0$, for GR, N5 and N1 respectively. The corresponding numbers from the L62 runs are much smaller (even with a larger box size), and so we do not quote them here -- this reflects the fact that the formation of disc galaxies is sensitive to the simulation resolution. 
From these numbers we do not observe any statistically significant trend of the impact of modified gravity. This is different from the case of $f(R)$ gravity \citep{Arnold:2019vpg}, which in the case of $|f_{R0}|=10^{-5}$ (F5) was found to produce significantly fewer disc galaxies than in GR. A possible explanation is the effect of modified gravity to enhance galaxy mergers, which makes it harder for disc galaxies to survive. In $f(R)$ gravity, we note a strong difference in halo abundance from GR \citep[see, e.g.,][]{Shi:2015aya}, which indicates that halo formation is strongly affected by the fifth force. This is, however, not the case in the nDGP models studied here \citep{Hernandez-Aguayo:2020b}, implying a weaker effect of the fifth force on the halo formation (by mergers and accretions). We shall leave a more careful analysis of the halo merger history to a future work.

%---------------------------------------------------------------
\section{Results}
\label{sec:Results}
%---------------------------------------------------------------
%---------------------------------------------------------------
\subsection{Clustering of matter components}
\begin{figure*}
 \centering
\includegraphics[width=1\textwidth]{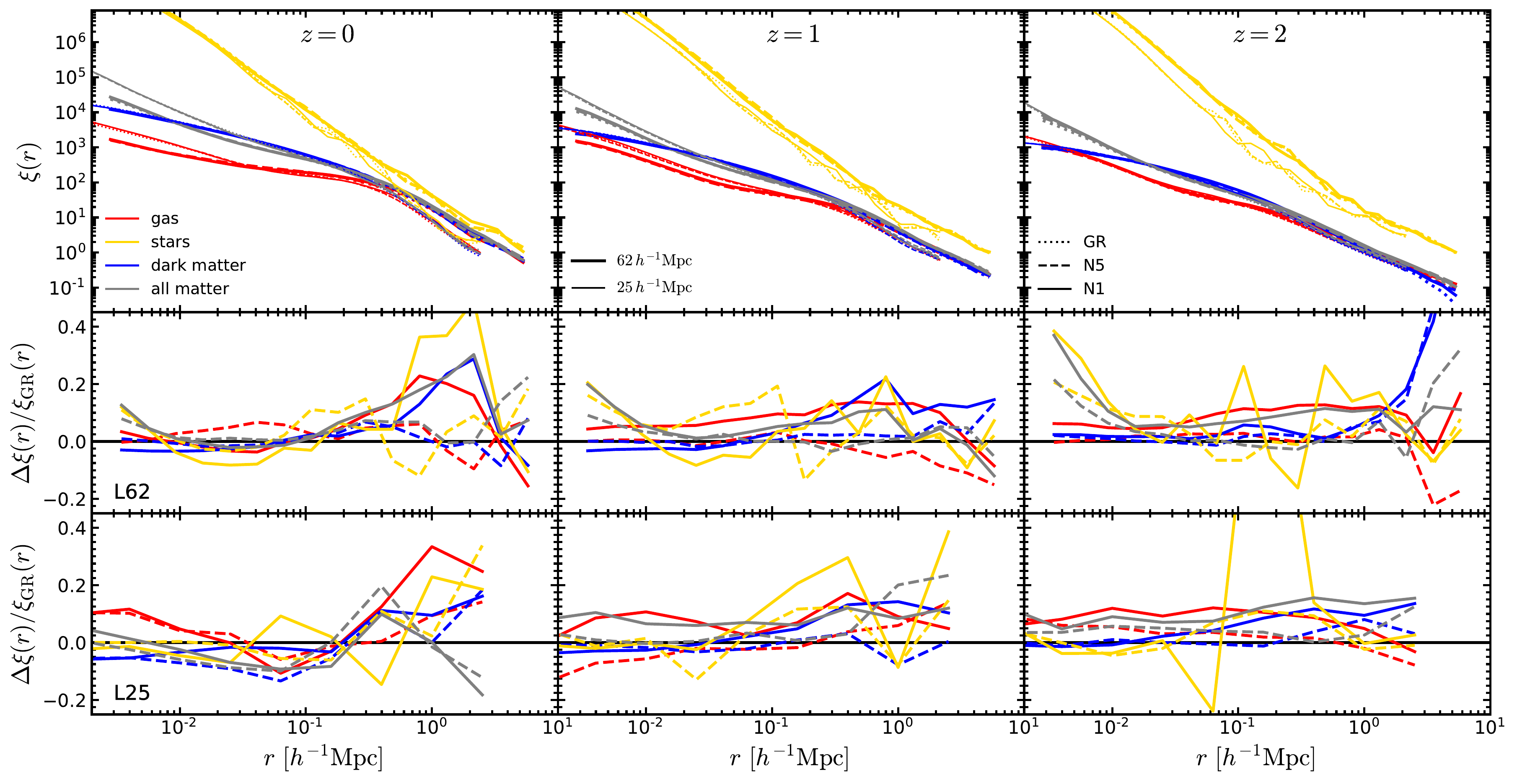}
\caption{The same as Fig.~\ref{fig:Pk_all} but for the correlation functions of the different matter components.}
\label{fig:xi_all}
\end{figure*}

%---------------------------------------------------------------
The measured power spectra and correlation functions for all types of matter components in our simulations are displayed in the upper panels of Figs.~\ref{fig:Pk_all} and \ref{fig:xi_all} at redshifts $z=0,1$ and $2$. We show results for the clustering of all gas (including both hot and cold components; red lines), stars (yellow), dark matter (blue), and the combination of all components (grey). The middle and lower rows of Fig.~\ref{fig:Pk_all} and Fig.~\ref{fig:xi_all} show the relative differences of the clustering measurements from the nDGP simulations with respect to GR for the L62 and L25 boxes, respectively. For the power spectrum (upper panels of Fig.~\ref{fig:Pk_all}) we additionally show the linear theory dark matter power spectrum for comparison as the light grey solid line.

From the upper panels of Fig.~\ref{fig:Pk_all} we note that the power spectra of different matter components have different behaviour and amplitudes, with stars being more clustered than the other types of matter irrespective of the gravity model. The clustering of dark matter and the total matter distribution show almost the same amplitude and follow the linear theory prediction on large scales through cosmic time. The power spectrum of gas displays a decrease in amplitude at the present time on small and intermediate scales; this behaviour is due to strong feedback effects that suppress galaxy formation at late times \citep{Springel:2017tpz}.

We can also see the impact of the simulation particle resolution on the matter power spectrum by comparing the thick and thin lines in Fig.~\ref{fig:Pk_all}. The main differences are the lack of large-scale modes in the L25 box, while the results of the L62 boxes are affected by the relatively low resolution on small scales. The most affected component due to resolution effects is the stars, which display a consistently higher amplitude at $z=2$ for the L62 box compare to the L25 box; However, this difference decreases at low redshifts. This is because stars in our low-resolution box (L62) tend to occupy higher-mass haloes than in the higher-resolution case and these haloes are more strongly biased. Also, as we will see later, the star formation rate is different between the L62 and L25 runs, which can also have an impact on the spatial distribution and clustering of stars. However, the GR results from both simulation boxes (thick and thin dashed lines, respectively) are consistent with the IllustrisTNG findings at different resolutions reported by \citet{Springel:2017tpz}. The same discussion on resolution effects on power spectrum applies to nDGP models as well.

%--------- Figure --------------

The interplay between baryons and modified gravity can be seen in the middle and bottom panels of Fig.~\ref{fig:Pk_all} for the L62 and L25 boxes, respectively. For dark matter, we can observe an enhancement on large scales $(k\lesssim 1\hMpc)$ due to the fifth force, leading to a maximum difference of $\sim 5\%$ and $\sim 20\% - 25\%$ for the N5 (dashed lines) and N1 (solid lines) models, respectively. This enhancement is consistent with results found with DM-only simulations \citep[see, e.g.,][]{Winther:2015wla}. On small scales, we note a suppression due to the Vainshtein screening mechanism. At $z\leq1$, there is a decrease of matter clustering in nDGP compared to GR at $k/(h{\rm Mpc}^{-1})\gtrsim10$. This may be partly due to the gravitational effect of gas and stars, but as we will see below, even in the DMO L62 simulations we see a similar suppression of dark matter power spectrum on these scales, which is a new feature only seen at high resolution.

The gas power spectrum (red lines) follows the same behaviour as dark matter on large scales at all redshifts. At early times ($z\geq1$), we observe that the gas power spectrum is less suppressed than dark matter on small scales; this is due to haloes that were able to accrete more gas from their surroundings, leading to a higher concentration of gas inside haloes, particularly for the N1 model. At the present time, the gas power spectrum is suppressed by $\sim 20-25\%$ for N1 in both boxes, while for N5 this effect is only observed for the L25 box on intermediate scales $5 < k/[\hMpc] < 40$. This is caused by stellar and AGN feedback that expels gas from inside the haloes, and it suggests that the feedback effects are stronger in the N1 model than GR.

The clustering of stars (yellow lines) is less affected by modified gravity than gas and dark matter, for which we find differences of $\lesssim 5\%$ for both N5 and N1 models and all redshifts, except at $z=2$ for the L62 box where the clustering of stars shows an increased clustering of $>10\%$ for both nDGP models (we caution about the L62 results regarding stars, given that the star clustering is strongly resolution dependent). The small difference between star clustering in the different models is the result of the Vainshtein screening mechanism inside haloes. Note that this is conceptually different to the behaviour of the stellar power spectrum in $f(R)$ gravity, where the clustering of stars is strongly influenced by the MG model even at $z=2$ \citep{Arnold:2019vpg}.
%--------- Figure --------------
\begin{figure*}
 \centering
\includegraphics[width=1\textwidth]{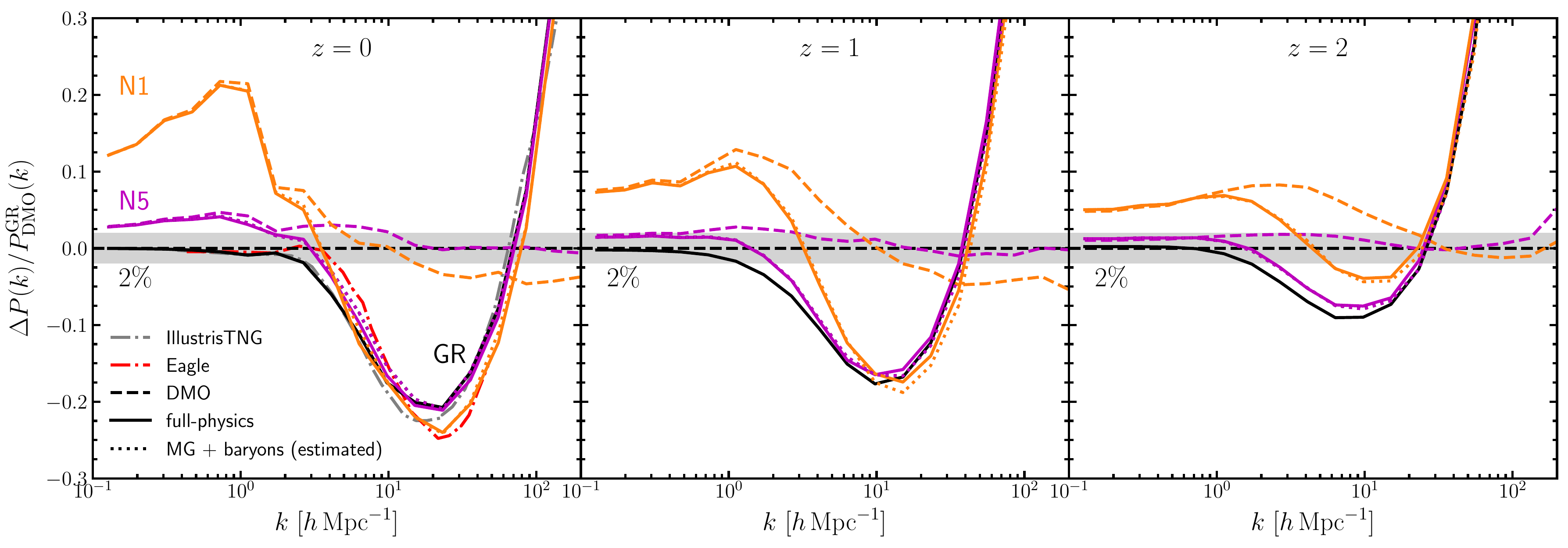}
\caption{The relative differences of the total matter power spectra from the full-physics (solid lines) and DM-only (dashed lines) L62 runs with respect to the matter power spectrum of the DM-only GR runs at $z=0$ ({\it left panel}), $z=1$ ({\it middle panel}) and $z=2$ ({\it right panel}). The grey and red dash-dotted lines show the impacts of baryons on the total matter power spectrum in the IllustrisTNG \citep{Springel:2017tpz} and Eagle \citep{Hellwing:2016ucy} simulations at $z=0$. Solid lines illustrate results from full-physics runs, dashed lines their DM-only counterparts. Dotted lines show an estimate for the combination of baryonic feedback and modified gravity effects, obtained by adding the relative differences of the nDGP DMO results to the GR full-physics run, cf.~Eq.~\eqref{eq:dp_model}. The different colours represent different gravity models as indicated in the {\it left panel} (black for GR, magenta for N5 and orange for N1). The light grey shaded region indicates a relative difference of 2 per cent.}
\label{fig:diff_all}
\end{figure*}

In Fig.~\ref{fig:xi_all} we show the correlation functions results for the same matter components and redshifts shown in Fig.~\ref{fig:Pk_all}. We find consistent trends with the power spectrum results discussed above.
We note that the gas correlation function starts to deviate from the dark matter and total matter correlation functions on scales $r\sim 0.2\Mpch$ at $z=2$. This evolves with time, when at $z=0$ the gas is much less clustered than dark matter on small scales. Again, the overall behaviour of the correlation functions and the resolution effects due to different box size is the same as the found for power spectrum. The relative differences between the nDGP models and GR are consistent with the power spectrum findings (see middle and bottom panels of Fig.~\ref{fig:xi_all}). In this case, we find that the correlation function of stars is noise dominated, making it difficult to observe a consistent difference between both nDGP models and GR as seen from the power spectrum.

Note that the clustering measured from the L25 simulation boxes can be strongly affected by sample variance, and so these simulations are not ideal to study the clustering differences between models of gravity. However, we show these results for completeness.
%---------------------------------------------------------------
\subsection{Impact of baryonic physics on the clustering of matter}
\begin{figure*}
 \centering
\includegraphics[width=1\textwidth]{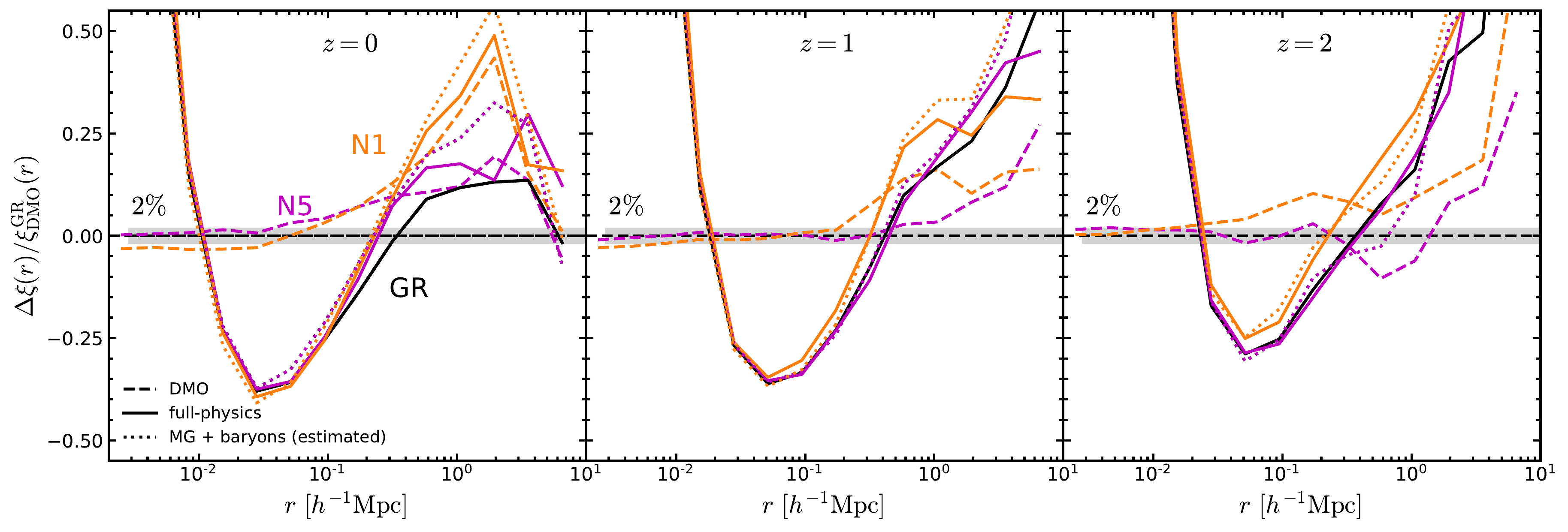}
\caption{The same as Fig.~\ref{fig:diff_all} but for the correlation function.}
\label{fig:diff_all_xi}
\end{figure*}

%---------------------------------------------------------------
In the first series of the SHYBONE simulations, \citet{Arnold:2019vpg} presented results on the degeneracy between the baryonic processes and modified gravity using the Hu-Sawicki $f(R)$ model \citep{Hu:2007nk}. Here, we are able to extend these findings to the nDGP braneworld model. Recall that we only produced DM-only runs for the L62 box (see Table \ref{tab:sims} for details), hence the results shown in Figs.~\ref{fig:diff_all} and \ref{fig:diff_all_xi} correspond to the large box of the SHYBONE-nDGP simulations.

Fig.~\ref{fig:diff_all} shows the relative differences between the full-physics power spectra of all matter in the three gravity models with respect to the dark (or equivalently all) matter power spectrum of the DM-only GR simulation at $z=0$, $1$ and $2$. We also show the predictions from the DM-only nDGP simulations (dashed lines) at the same redshifts. On large scales ($k<2\hMpc$) we find a consistent enhancement of the DM-only power spectrum of the N5 and N1 models (dashed lines) with the dark matter component of the full-physics runs (see blue lines in the middle panels of Fig.~\ref{fig:Pk_all})

At $z=0$, we can see a suppression in the matter power spectrum of $\sim 20\%$ for GR and N5 models at scales $k\sim 20\hMpc$, while for N1 the power is suppressed by $\sim 25\%$. This suppression becomes smaller with increasing redshift; as shown in \citet{Springel:2017tpz} and \citet{Arnold:2019vpg}, one should expect a negligible baryonic effect on intermediate and large scales at $z>3$. 
For comparison, we also show the results from the IllustrisTNG \citep{Springel:2017tpz} and Eagle \citep{Hellwing:2016ucy} simulations at $z=0$, noting the good agreement with our GR results. 
The significant enhancement of the matter power spectrum for $k>40\hMpc$ is consistent with the IllustrisTNG result, but in our case in the highly resolution-affected regime.

The dotted lines display the estimated effect of baryonic feedback from the GR full-physics simulation added to the predictions from the DM-only nDGP simulations:
\begin{equation}\label{eq:dp_model}
    \frac{\Delta P(k)}{P_{\rm DMO}^{\rm GR}(k)} = \left[\frac{P^{\rm nDGP}_{\rm DMO}(k)}{P^{\rm GR}_{\rm DMO}(k)}-1\right] + \left[\frac{P^{\rm GR}_{\rm full 
    -physics}(k)}{P^{\rm GR}_{\rm DMO}(k)}-1\right].
\end{equation}
The idea is that the impacts of baryonic physics and modified gravity can be relatively clearly separated and their back-reaction effects on each other are negligible \citep{Arnold:2019vpg}. This figure shows this simple model is accurate enough to reproduce the full-physics results in nDGP simulations: comparing the magenta and orange dotted lines with their solid line counterparts, we can see the agreement is generally at percent level at all scales up to $k\simeq100\hMpc$. 
On scales $k\lesssim1\hMpc$, the effect of baryons is negligible in both DGP models and at all redshifts, showing that the relative differences are dominated by the modified gravity effect reaching a maximum value with the same amplitude as the DM-only simulations. 

%--------- Figure --------------

In Fig.~\ref{fig:diff_all_xi} we explore the impact of the baryonic feedback effects on the correlation functions at the same redshifts ($z=0$, $1$ and $2$). We find several differences from the power spectrum results. First, the clustering in configuration space also shows a suppression on small scales, but the difference is larger than that found for the power spectrum. In this case, the relative change is $\sim40$ per cent at $z=0$,  decreasing with redshift to $25$ per cent at $z=2$ (recall that we find a maximum difference of $25$ per cent at $z=0$ and about $10$ per cent at $z=2$ in the power spectrum). We observe the same enhancement on very small scales ($r<10\,h^{-1}{\rm kpc}$) as the power spectrum. But the clustering of matter shows an increase of $10$ per cent at $r=1\Mpch$ for the GR simulations at $z=0$, which becomes even larger at earlier times. 

In the nDGP models, the full-physics matter correlation functions follow the same trend as in GR, but the enhancement on large scales is bigger for the N1 model which reaches a maximum difference of 50 per cent at the present epoch, followed by the N5 model which presents a similar value to the GR case. Nevertheless, due to the relatively small box size of our simulations and the comparatively more noisy measurement of the correlation function at large radii, we do not observe the constant enhancement at large-scales as shown in the power spectrum.

From Fig.~\ref{fig:diff_all_xi} we also note that estimating the feedback impact by adding the GR full-physics effect to the DM-only difference in modified gravity models (dotted lines), does lead to good agreement with the nDGP full-physics results as seen in the power spectrum, especially on scales $<0.5\Mpch$. This approximation works less well on scales beyond $0.5\Mpch$, where we find larger differences in the relative values by comparing the solid and dotted lines in all panels of Fig.~\ref{fig:diff_all_xi}. We caution again that the differences on large-scales between the full-physics and the estimated impact of baryons in the nDGP models (solid and dotted lines in Fig.~\ref{fig:diff_all_xi}) could be due to the limited size of our simulation box.

The dark matter clustering measured from the DM-only simulations (dashed lines) of the nDGP models shows a similar trend to the dark matter component of the full-physics run (see Fig.~\ref{fig:xi_all}). We find that on very small scales, the clustering of the dark matter is very close to GR, but the N1 model displays a slight suppression at $z\leq1$, which is consistent to the small scale suppression of the matter power spectra in Fig.~\ref{fig:diff_all} (see also the discussion of Fig.~\ref{fig:Pk_all} above). On scales $>100\,h^{-1}{\rm kpc}$ the effect of modified gravity increases the amplitude of the clustering at all redshifts. This is also consistent with the power spectrum results presented in Fig.~\ref{fig:diff_all}.
%--------- Figure --------------
\begin{figure*}
 \centering
\includegraphics[width=0.9\textwidth]{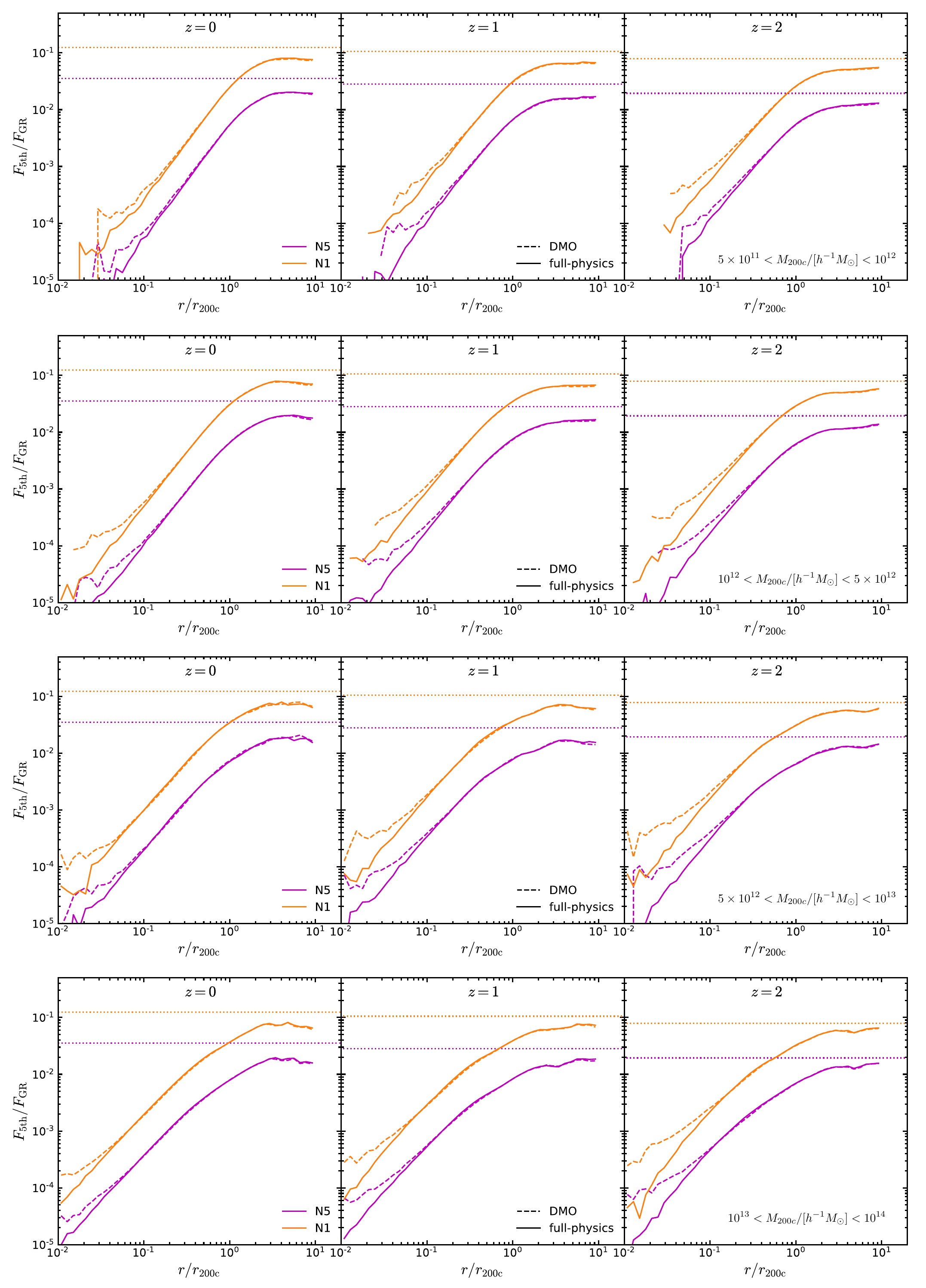}
\caption{Radial profiles of the fifth-to-Newtonian force ratio of dark matter haloes in the full-physics (solid lines) and DM-only (dashed lines) nDGP-L62 simulations (magenta for N5 and orange for N1) at $z=0$ ({\it left panels}), $z=1$ ({\it middle panels}) and $z=2$ ({\it right panels}). We show the results for four mass bins: $M_{200c}=(5\times10^{11}-10^{12})\Msh$, $(10^{12}-5\times10^{12})\Msh$, $(5\times10^{12}-10^{13})\Msh$ and $(10^{13}-10^{14})\Msh$ (from {\it top} to {\it bottom}). The horizontal dotted lines show the value, $F_{\rm 5th}/F_{\rm GR} = 1/3\beta(a)$, for each nDGP model and redshift.}
\label{fig:F_profile}
\end{figure*}
%--------- Figure --------------
\begin{figure*}
 \centering
\includegraphics[width=0.9\textwidth]{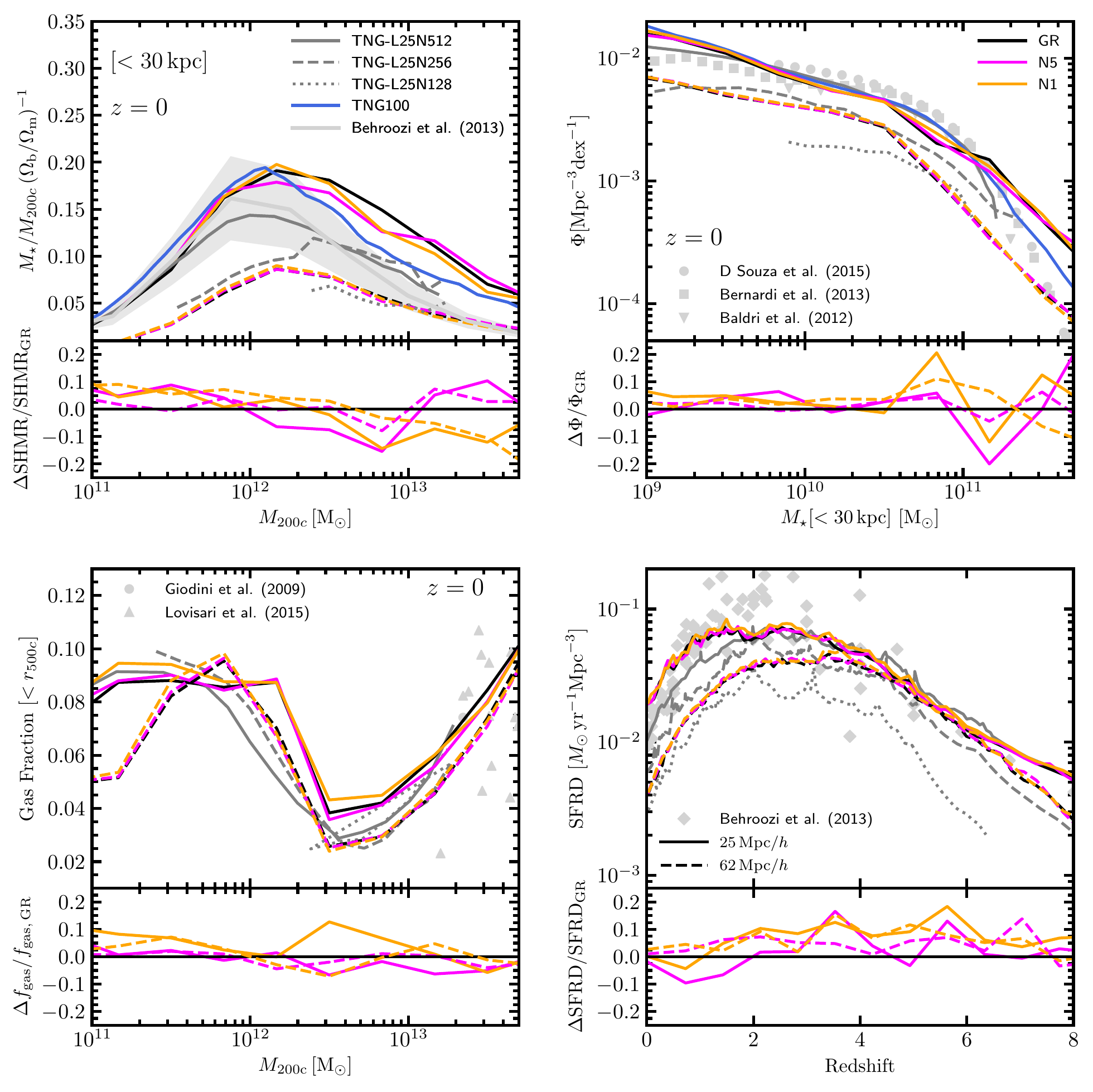}
\caption{Stellar and gaseous galaxy properties at $z=0$ (unless otherwise stated). {\it Upper left:} Stellar-to-halo mass ratio as a function of halo mass. The stellar mass is measured within 30 kpc from the halo centre. {\it Upper right:} Galaxy stellar mass function measured within 30 kpc from the centre of the halo. {\it Bottom left:} The halo gas fraction within $r_{500c}$ as a function of the total halo mass. {\it Bottom right:} Star-formation rate density as a function of redshift. Different colours represent different gravity models as specified in the legend. Solid colour lines show results from our L25 simulations while dashed coloured lines are from our L62 boxes. All lower subpanels show the relative differences between the galaxy properties of nDGP and GR models. In all panels we compare our results with the IllustrisTNG results at three different resolutions \citep{Pillepich:2017jle}: TNG-L25N512 (solid grey curves), TNG-L25N256 (dashed grey curves) and TNG-L25N128 (dotted grey curves). The blue solid line in the upper panels shows the results from the TNG100 simulation \citep{Pillepich:2017fcc}. Light-grey symbols represent observational measurements from: \citet{Behroozi:2013iw} \citet{Baldry:2011fj}, \citet{Bernardi:2013mqa}, \citet{DSouza:2015D}, \citet{Giodini:2009qf} and \citet{Lovisari:2015pka}.}
\label{fig:galaxy_properties}
\end{figure*}

%---------------------------------------------------------------
\subsection{Fifth force profiles}
%---------------------------------------------------------------
We can also explore the interplay between modified gravity and baryonic effects by looking at the fifth force profiles of dark matter haloes. The median of the fifth-to-Newtonian force ratio profiles in our nDGP-L62 (full-physics and DM-only) runs using four mass bins: $M_{200c}=(5\times10^{11}-10^{12})\Msh$, $(10^{12}-5\times10^{12})\Msh$, $(5\times10^{12}-10^{13})\Msh$ and $(10^{13}-10^{14})\Msh$ at $z=0$, $1$ and $2$ as a function of $r/r_{200c}$ (with $r$ the distance from the halo centre) is shown in Fig.~\ref{fig:F_profile}. 

We observe the suppression of the fifth force inside the haloes due to the Vainshtein screening mechanism. Far from the centre, the fifth force approaches the value
\begin{equation}
    F_{\rm 5th} = \frac{1}{3\beta}F_{\rm GR},
\end{equation}
as expected [cf.~Eq.~\eqref{eq:fifth-force-ratio-asymptotic}]. This value is showed as the dotted lines in Fig.~\ref{fig:F_profile}. There is a strong suppression of the fifth force in the inner regions of the haloes ($r < 0.1\,r_{200c}$) at all three redshifts, due to higher densities close to the centres of the haloes. 
This can be seen from the following solution to ${\rm d\varphi}/{\rm d}r$ for a general spherical density profile $\rho(r)$, which can be obtained from Eq.~\eqref{eq:dgp} or \eqref{eq:dgp_eqn}:
\begin{equation}\label{eq:F5_dgp}
    \frac{{\rm d}\varphi}{{\rm d}r} = \frac{\sqrt{1+\frac{64\pi Gr^2_{\rm c}}{27\beta^2c^2}\bar{\rho}(<r)}-1}{\frac{16\pi Gr_{\rm c}^2}{9\beta c^2}\bar{\rho}(<r)}g_{\rm N}(r),
\end{equation}
where $\bar{\rho}(<r)$ denotes the mean matter density within radius $r$ from the halo centre, and we have again set $a=1$ for simplicity. In high-density regions, the second term in the square root of the numerator dominates over the first term, so that the fifth-to-Newtonian force ratio decays as $\bar{\rho}(<r)^{-1/2}$. For the same reason, in the inner regions we see that the fifth force is more suppressed in the full-physics than in the DM-only runs, since gas and stars condensate at halo/galactic centres, increasing $\bar{\rho}(<r)$ there.

We also find that the force ratio profiles in Fig.~\ref{fig:F_profile} do not show a noticeable dependence on the halo mass, consistent with previous findings of DMO simulations \citep{Winther:2015wla}. This behaviour can also be explained using Eq.~\eqref{eq:F5_dgp}, which indicates that the force ratio only depends on $\bar{\rho}(<r)$. At $r=r_{200c}$, we have that $\bar{\rho}(<r)$ is equal to 200 times the critical density, independent of halo mass; a weak dependence on halo mass is introduced due to the different density profiles (concentrations), but the effect is small.

We note that the median force profiles do not reach the expected linear-theory value (dotted horizontal lines in Fig.~\ref{fig:F_profile}) at $r\sim10r_{200c}$, especially at $z=0$. However, recall that the linear value is expected to be reached only at $r \gg r_V$, cf.~Eq.~\eqref{eq:fifth-force-ratio-asymptotic}. It can be calculated that the Vainshtein radius (see Eq.~\eqref{eq:r_V}) of haloes with radius $r=r_{200c}$ at $z=0$ is $r_V \sim 3.4\times r_{200c}$ and $r_V \sim 2.6\times r_{200c}$ for N5 and N1, respectively, which is of the same order of $10r_{200c}$.
Furthermore, haloes are not isolated objects, but are likely to be surrounded by neighbouring haloes or filaments, which means that going outwards we should not expect the fifth-force-to-Newtonian-gravity ratio profile to monotonically asymptote to the linear-theory prediction --- actually, the curves in Fig.~\ref{fig:F_profile} tend to peak and start to decay again (or flatten) at $r\simeq5\times r_{200c}$, which could be a signature that something (possibly other haloes) are being encountered. We have also checked (not shown here) that there is a substantial scatter across individual haloes: inside some haloes the force profile can get very close to the linear-theory values at $r\simeq5\times r_{200c}$, while for others it is significantly lower --- therefore, the net effect is that the median value does not reach the background value. In any case, our spherical tophat overdensity test shows that the numerical solution of the scalar field (and thus the fifth force) is in very good agreement with the analytical prediction for isolated objects (see Appendix~\ref{sec:code_test} for details).

%---------------------------------------------------------------
\subsection{Galaxy properties in braneworld models}
%---------------------------------------------------------------
In Fig.~\ref{fig:galaxy_properties} we show the results on galaxy demographics of the SHYBONE-nDGP simulations. Recall that we did not re-tune the IllustrisTNG galaxy formation model to match observations in modified gravity, hence we use the same hydrodynamical model for all gravity models. In all panels of Fig.~\ref{fig:galaxy_properties} we compare our results with the TNG-L25 boxes reported by \citet{Pillepich:2017jle}. These TNG-L25 test simulations were run using a box with size of $L=25\Mpch$ and three mass resolutions: $2\times512^3$ (high-resolution, TNG-L25N512), $2\times256^3$ (medium-resolution, TNG-L25N256) and $2\times128^3$ (low-resolution, TNG-L25N128) dark matter and gas elements. Depending on the panel and galaxy property, we show observational data of the stellar-to-halo mass ratio and star formation rate density from \citet{Behroozi:2013iw}, galaxy stellar mass from \citet{Baldry:2011fj}, \citet{Bernardi:2013mqa} and \citet{DSouza:2015D}; and gas fractions from \citet{Giodini:2009qf} and \citet{Lovisari:2015pka}. 
We do not expect our L62 results to match the obervational data, since the TNG model was tuned for the TNG-L25N512 test simulations using the galaxy properties mentioned above \citep[as well as the black hole mass - stellar mass relation and the galaxy size at redshift $z=0$;][]{Pillepich:2017jle} and it has been demonstrated that the stellar properties of galaxies depend on the simulation resolution in the TNG model \citep[see Appendix A of][for details]{Pillepich:2017jle,Pillepich:2017fcc}. 

The upper left panel of Fig.~\ref{fig:galaxy_properties} shows the stellar-to-halo mass ratio multiplied by the inverse of the baryon fraction $(\Omega_{\rm b}/\Omega_{\rm m})$, as a function of the total host halo mass $(M_{200c})$ for our six full-physics simulations at $z=0$. The stellar mass was measured within $30$ kpc from the halo centre. First, we note that the L62 simulations (dashed coloured lines) are in good agreement with the TNG-L25N128 (grey dotted line) run at the high-mass end, but is lower than the TNG-L25N256 (grey dashed line) and TNG-L25N512 (grey solid line) results, which is as expected given the resolutions of these runs.
Our L25 runs (solid coloured lines) predict higher stellar mass fractions than the TNG-L25N512 (grey solid line), despite the fact that these simulations have the same resolution. The differences are due to the combination of the different initial condition realisations we used to run our simulations (i.e., cosmic variance) and the small number of high mass objects in the small boxes. Note that the final TNG100 run \citep[blue solid line;][]{Pillepich:2017fcc} also has disagreements with the TNG-L25N512 test run, for the same reason. Nevertheless, our L25 predictions are in good agreement with observational estimates (light grey area) and with TNG100 at $M_{200c}\lesssim2\times10^{12}M_\odot$. We find that the modified gravity effects induce a $\sim 10\%$ change with respect to GR for small haloes in both nDGP models, and the relative differences in the L62 and L25 boxes are consistent with each other over most of the mass range. 
However, the nDGP-L25 runs produce a higher model difference in the stellar mass fraction for haloes with mass $M_{200c}\sim10^{13}M_\odot$, than their L62 counterparts. However, we again caution here that at this mass the simulations, especially L25, may suffer from cosmic variance.

The galaxy stellar mass functions (GSMF) measured within 30 kpc from halo centres are shown in the upper right panel of Fig.~\ref{fig:galaxy_properties}. Our L62 results are consistent with the TNG-L25N256 simulations, while the L25 GSMFs are in excellent agreement with the TNG-L25N512 and TNG100 simulations at $M_\star\lesssim2\times10^{11}M_\odot$.
For both nDGP models, both L25 and L62 display small differences, $\sim5\%$, with respect to the GR counterparts, at $M_\star\lesssim4\times10^{10}M_\odot$. At even higher stellar masses, the relative difference curves are noisier and the agreement between L25 and TNG100 is poorer, due to the small box size and due to the low number of galaxies in the high-mass end. Overall, we conclude that the GSMF is not significantly altered by either of the nDGP models studied here.

The lower left panel of Fig.~\ref{fig:galaxy_properties} displays the galaxy gas fractions within $r_{500c}$ as a function of halo mass. We can see that, considering the scatters of observational data (grey circles and triangles), both sets of SHYBONE-nDGP simulations are in good overall agreement with the three TNG-L25 tests, except for the L62 run for masses $M_{200c}\lesssim 4\times10^{11}M_\odot$ due to the limited mass resolution. 
In addition, for both L62 and L25, the relative differences between the nDGP models and GR are consistent with each other in almost the entire halo mass range. The small difference in the relative difference lines between L62 and L25 is again likely to be noise, and this plot does not point to a strong effect of modified gravity.

We present the star-formation rate densities (SFRD) as a function of time in the lower right panel of Fig.~\ref{fig:galaxy_properties}. We confirm the findings of \citet{Pillepich:2017jle} and \citet{Arnold:2019vpg} that the SFRDs are resolution dependent, particularly at low redshift. 
Our L62 simulations show a higher star formation rate than the TNG-L25N256 and TNG-L25N128 at high redshift ($z>4$), while at lower redshifts ($z<3$) the curves fall between the low- and medium-resolution TNG-L25 boxes, as expected from their resolution. The L25 boxes are in excellent agreement with the TNG-L25N512 run, but display a higher SFRD at lower redshifts ($z<2$), which is nevertheless still in agreement with the observational data (light grey symbols). Note that the nDGP-L62 relative differences are in good agreement with our high-resolution runs (L25) displaying a maximum relative change of $\sim15\%$ with respect to GR. However, at the present time, the SFRD in both nDGP models match the GR predictions within a $3\%$ margin.

Due to the small (compared to scatters in observational data) differences that the nDGP models induce in the properties of galaxies, we arrive at the same conclusion as for $f(R)$ gravity in \citet{Arnold:2019vpg}: a re-tuning of the TNG model for nDGP gravity is not necessary and this allows us to study different gravity models using the same prescription for galaxy formation. This also indicates that the global galaxy properties shown in Fig.~\ref{fig:galaxy_properties} cannot be used to distinguish between the different gravity models, at least not with the current level of observational and simulation uncertainties.

The small impact of modified gravity on the global galactic and gas properties can be partly explained by the Vainshtein screening mechanism which, as we have seen, effectively suppresses the fifth force inside haloes, cf.~Fig.~\ref{fig:F_profile}. However, we do see nDGP effects at the level of about $5-15\%$ in Fig.~\ref{fig:galaxy_properties}: this is because galaxy formation is a complicated process that is not confined to the inner regions of haloes, but the recycle of gas actually involves regions in the outer parts of, or even outside, haloes, where Vainshtein screening is less effective.

%---------------------------------------------------------------
\section{Summary and conclusions}
\label{sec:conc}
%---------------------------------------------------------------
We introduced a new set of galaxy formation simulations in the DGP braneworld model. In order to run these simulations, we extended the modified version of the {\sc Arepo} code for $f(R)$ gravity presented by \citet{Arnold:2019vpg}, so that it can be used to simulate structure formation in the DGP model. We performed a series of tests to verify that the new code gives reliable results. We also compared the matter power spectrum predicted by the new code with predictions by the {\sc ecosmog-V} code \citep{Li:2013nua}, finding excellent agreement on all scales.

This implementation, together with the IllustrisTNG galaxy formation model desribed by \citet{Pillepich:2017jle}, makes the new set of ``full-physics'' hydrodynamical simulations in DGP gravity possible. The simulations we used in our analysis employ $2\times512^3$ dark matter particles and gas cells. We studied two cosmological volumes: a large-volume simulation with box size $L=62\Mpch$ (L62 runs), and a  small-volume with box size $L=25\Mpch$ (L25 runs). For each set, the simulations cover three gravity models -- GR, N5 and N1. These are supplemented by DM-only simulations for the same models and using the same specifications as the L62 full-physics runs. We have saved 100 snapshots per run, which contain all the particle data and group catalogues generated using {\sc subfind}. In a future paper, we shall build merger trees from the 100 DM-only snapshots to use in semi-analytical models of galaxy formation.

We studied the real-space clustering in Fourier and configuration space of stars, gas, dark matter and the total matter distribution. The clustering of dark matter in the full-physics nDGP simulations displays an enhancement compared to its GR counterpart on large scales ($k<4\hMpc$ for the power spectrum), consistent with previous findings from DM-only simulations \citep{Winther:2015wla}. The clustering of gas and the total matter distribution follows a similar trend to the dark matter on almost all scales, and the clustering of the stellar content seems to be less affected by changes in the gravity model in Fourier space. 

We find that the interplay between baryonic feedback processes and modified gravity is complex. However, the impact of baryons on the clustering of matter has a similar impact in all gravity models, with a suppression of $\sim 25$ per cent in the power spectrum and up to $\sim 40$ per cent in the correlation function at the present time. In particular, the impacts of baryons and modified gravity on the matter power spectrum -- and to a similar extent on the correlation function -- can be modelled additively by summing up their changes to the GR dark matter power spectrum, with a percent-level accuracy, cf. Figs.~\ref{fig:diff_all} and \ref{fig:diff_all_xi}.

The fifth force to normal gravity ratio in dark matter haloes, $F_{\rm 5th}/F_{\rm GR}$, is also affected by baryons. We found a suppression of the ratio due to higher densities in the inner regions (close to the centre) of the haloes in full-physics runs relative to DMO. Additionally, we showed that the force profiles have only a weak dependence on halo mass, confirming the findings of DM-only simulations presented by  \citet{Winther:2015wla}, and note that this is a feature that is expected for Vainshtein screening.

The stellar and gaseous properties of galaxies are only mildly affected by the modifications to gravity in the nDGP models, mirroring the results found by \citet{Arnold:2019vpg} for the case of $f(R)$ gravity. The differences induced by the nDGP model are nevertheless even smaller than those caused by $f(R)$ gravity. Therefore, we conclude that given the current size of uncertainties in the relevant galactic observables, there is no need to ru-tune the baryonic physics model for these modified gravity models.

The SHYBONE simulations (both for $f(R)$ gravity and nDGP models) aim to assist future galaxy surveys by making predictions for the small-scale galaxy clustering and stellar properties in galaxies. We plan to continue improving our MG simulation techniques, for example, to simulate Milky-Way and Local-Group like systems through zoom techniques. These simulations will help us to understand the impact of modified gravity on small cosmological scales and the more complex of the astrophysical processes. In a series of follow-up works based on the new simulations, we will investigate such impacts on a range of observational quantities in greater depth.

%---------------------------------------------------------------
\section*{Acknowledgements}
%---------------------------------------------------------------
CH-A acknowledges support from the Mexican National Council of Science and Technology (CONACyT) through grant No. 286513/438352 and from the Excellence Cluster ORIGINS which is funded by the Deutsche Forschungsgemeinschaft (DFG, German Research Foundation) under Germany's Excellence Strategy - EXC-2094 - 390783311. CA and BL are supported by the European Research Council through ERC Starting Grant ERC-StG-716532-PUNCA. All authors acknowledge support from the Science Technology Facilities Council (STFC) through ST/T000244/1 and ST/P000541/1. 
This work used the DiRAC@Durham facility managed by the Institute for Computational Cosmology on behalf of the STFC DiRAC HPC Facility (\url{www.dirac.ac.uk}). The equipment was funded by BEIS capital funding via STFC capital grants ST/K00042X/1, ST/P002293/1, ST/R002371/1 and ST/S002502/1, Durham University and STFC operations grant ST/R000832/1. DiRAC is part of the National e-Infrastructure.

%---------------------------------------------------------------
\section*{Data Availability}
%---------------------------------------------------------------
The data underlying this article will be shared on reasonable request to the corresponding authors.
%%%%%%%%%%%%%%%%%%%%%%%%%%%%%%%%%%%%%%%%%%%%%%%%%%



\bibliographystyle{mnras}
\bibliography{ref} % if your bibtex file is called example.bib

%%%%%%%%%%%%%%%%%%%%%%%%%%%%%%%%%%%%%%%%%%%%%%%%%%

%%%%%%%%%%%%%%%%% APPENDICES %%%%%%%%%%%%%%%%%%%%%

\appendix
%---------------------------------------------------------------
%\section{Appendix}
%---------------------------------------------------------------
%---------------------------------------------------------------
\section{Numerical methodology}
\label{sec:methods}
%---------------------------------------------------------------
N-body cosmological simulations have played an important role in the {study} of alternative gravity models, allowing to study the impact of modified gravity on the clustering of galaxies. Such simulations are necessary for the construction of synthetic galaxy catalogues. In this section we present the numerical methods used to implement the nDGP model into {\sc Arepo}. Combined with the  IllustrisTNG galaxy formation model, this allows us to run full hydrodynamical simulations in the nDGP model.

%---------------------------------------------------------------
\subsection{N-body algorithm}\label{sec:nbody}
%---------------------------------------------------------------
The equation of motion of the brane-bending mode, Eq.~\eqref{eq:phi_dgp}, can be written using the code units \citep[see][for details]{Weinberger:2019tbd} of {\sc Arepo} as
\begin{equation}
\nabla^2 \varphi + \frac{R_{\rm c}^2}{3\beta\,a^3} \left[ (\nabla^2\varphi)^2 - (\nabla_{i}\nabla_{j}\varphi)^2 \right] = \frac{8\pi\,G\,a^2}{3\beta}\delta\rho\,,
\label{eq:phi_code}
\end{equation}
where $c$ is the speed of light and $G$ the gravity constant in internal code units. In this equation we have introduced a new dimensionless quantity,
\begin{equation}
R_{\rm c} \equiv \frac{r_{\rm c}}{c} = \frac{1}{2H_0\sqrt{\Omega_{\rm rc}}}.
\end{equation}
Eq.~\eqref{eq:phi_code} can be expressed, after applying the operator-splitting trick \citep{Chan:2009ew}, as follows
\begin{equation}\label{eq:phiQS-code}
\left(1-w\right)\left(\nabla^2\varphi \right)^2 + \alpha \nabla ^2\varphi  - \Sigma  = 0,
\end{equation}
where
\begin{equation}
\alpha = \frac{3\beta a^3}{R_{\rm c}^2},
\end{equation}
\begin{equation}
\Sigma = \left(\nabla_{i }\nabla_{j }\varphi \right)^2 - w\left(\nabla ^2\varphi \right)^2 + \frac{\alpha}{3\beta}8\pi G a^2\delta\rho\,,
\end{equation}
and $w$ is a constant numerical factor which has to be chosen as $1/3$ for the numerical algorithm to converge. Eq. (\ref{eq:phiQS-code}) can be solved once to yield
\begin{eqnarray}
\nabla ^2\varphi  &=& \frac{1}{2(1-w)}\left[-\alpha \pm\sqrt{\alpha ^2+4(1-w)\Sigma }\right]\nonumber\\
&=& \frac{1}{2(1-w)}\left[-\alpha +\frac{\alpha }{|\alpha |}\sqrt{\alpha ^2+4(1-w)\Sigma}\right],\label{eq:phiQS-2code}
\end{eqnarray}
where in the second line we have specialised to the relevant branch of the solution \citep{Li:2013nua}.

The discrete version of the field derivatives are
\begin{eqnarray}
\nabla\varphi &=& \frac{1}{2h}\left(\varphi_{i+1,j,k}-\varphi_{i-1,j,k}\right),\\
\nabla^2\varphi &=& \frac{1}{h^2}\left(\varphi_{i+1,j,k}+\varphi_{i-1,j,k}-2\varphi_{i,j,k}\right),\\
\nabla_x\nabla_y\varphi &=& \frac{1}{4h^2}\Big(\varphi_{i+1,j+1,k}+\varphi_{i-1,j-1,k}-\varphi_{i+1,j-1,k}\nonumber\\
&& ~~~~~~~~~~-\varphi_{i-1,j+1,k}\Big),
\end{eqnarray}
where $h$ is the cell length and we have assumed one dimension for simplicity for $\nabla\varphi$ and $\nabla^2\varphi$.

Instead of solving the full modified Poisson equation, \eqref{eq:poisson_nDGP}, to obtain the total gravitational potential $\Phi$, we split the force calculation into two parts: (i) solving the standard Poisson equation to get the Newtonian potential $\Phi_{\rm N}$ and hence calculate the Newtonian force, and (ii) solving the scalar field $\varphi$ to obtain the fifth force. The Newtonian force is obtained from the standard gravity solver implemented in {\sc Arepo} \citep[see][for details]{Springel:2009aa,Weinberger:2019tbd}. 
To solve the scalar field equation, we implemented an AMR modified gravity solver in Arepo \citep[see][for details of the MG solver and our local time-stepping scheme]{Arnold:2019vpg}. Therefore, the code uses a Voronoi mesh for the hydrodynamics solver and an AMR grid to solve the MG scalar field. These two grids are independent. On one hand, the Voronoi mesh `adapts' itself so that the gas mass in the cells is roughly the same (within a factor of 2). In this way the Voronoi mesh has a higher spatial resolution in high density areas, e.g., inside galaxies \citep[see][for details]{Springel:2009aa}; also, the Voronoi mesh does not refine in a classical sense, as there is only one mesh `level', but the spatial resolution varies with position and time. On the other hand, the AMR grid is refined in high (total) density regions, but it uses a classical approach with multiple mesh levels which allow for multigrid acceleration. The modified gravity solver then considers all masses to build the AMR grid and compute the forces --- these include DM, gas (from the Vorronoi cells), stars and black holes.

The EOM for the brane-bending mode, Eq.~\eqref{eq:phiQS-2code}, can be written as an operator equation 
\begin{equation}\label{eq:operator_eqn}
\mathcal{L}^h(\varphi_{i,j,k})=0,
\end{equation}
with
\begin{align}
\mathcal{L}^h(\varphi_{i,j,k})&\equiv\nonumber\\ \frac{1}{h^2}\Big(&\varphi_{i+1,j,k}+\varphi_{i-1,j,k}+\varphi_{i,j+1,k}\nonumber\\
 +&\varphi_{i,j-1,k}+\varphi_{i,j,k+1}+\varphi_{i,j,k-1}-6\varphi_{i,j,k}\Big)\nonumber\\
 -&\frac{1}{2(1-w)}\left[-\alpha+\frac{\alpha}{|\alpha|}\sqrt{\alpha^2+4(1-w)\Sigma_{i,j,k}}\right]\label{eq:Lphi},
\end{align}
in which the superscript $^h$ is used to label the level of the mesh (or equivalently the size of the cell of that level), and we have defined
\begin{widetext}
\begin{eqnarray}
\Sigma_{i,j,k} &\equiv& \frac{1-w}{h^4}\bigg[\Big(\varphi_{i+1,j,k}+\varphi_{i-1,j,k}-2\varphi_{i,j,k}\Big)^2+\Big(\varphi_{i,j+1,k}+\varphi_{i,j-1,k}-2\varphi_{i,j,k}\Big)^2+\Big(\varphi_{i,j,k+1}+\varphi_{i,j,k-1}-2\varphi_{i,j,k}\Big)^2\bigg]\nonumber\\
&& - \frac{2}{h^4}w\left(\varphi_{i+1,j,k}+\varphi_{i-1,j,k}-2\varphi_{i,j,k}\right)\left(\varphi_{i,j+1,k}+\varphi_{i,j-1,k}-2\varphi_{i,j,k}\right)\nonumber\\
&& - \frac{2}{h^4}w\left(\varphi_{i+1,j,k}+\varphi_{i-1,j,k}-2\varphi_{i,j,k}\right)\left(\varphi_{i,j,k+1}+\varphi_{i,j,k-1}-2\varphi_{i,j,k}\right)\nonumber\\
&& - \frac{2}{h^4}w\left(\varphi_{i,j+1,k}+\varphi_{i,j-1,k}-2\varphi_{i,j,k}\right)\left(\varphi_{i,j,k+1}+\varphi_{i,j,k-1}-2\varphi_{i,j,k}\right)\nonumber\\
&& + \frac{1}{8h^4}\Big(\varphi_{i+1,j+1,k}+\varphi_{i-1,j-1,k}-\varphi_{i+1,j-1,k}-\varphi_{i-1,j+1,k}\Big)^2\nonumber\\
&& + \frac{1}{8h^4}\Big(\varphi_{i+1,j,k+1}+\varphi_{i-1,j,k-1}-\varphi_{i+1,j,k-1}-\varphi_{i-1,j,k+1}\Big)^2\nonumber\\
&& + \frac{1}{8h^4}\Big(\varphi_{i,j+1,k+1}+\varphi_{i,j-1,k-1}-\varphi_{i,j+1,k-1}-\varphi_{i,j-1,k+1}\Big)^2 + \frac{\alpha}{3\beta}8\pi G a^2\delta\rho_{i,j,k}.
\end{eqnarray}
\end{widetext}
and
\begin{equation}
\delta\rho_{i,j,k} = \frac{m_{i,j,k}}{h^3} - \bar{\rho}(a)\,,
\end{equation}
where $m_{i,j,k}$ is the mass assigned to cell $(i,j,k)$ using a cloud-in-cell scheme, and $\bar{\rho}(a) = \bar{\rho}_0 / a^3$ is the mean physical matter density as a function of the scale factor. 

This equation can be solved by using the multigrid relaxation method, for which the code iterates to update the value of $\varphi_{i,j,k}$ in all cells, and at each iteration the field values changes as
\begin{equation}
\varphi^{h,{\rm new}}_{i,j,k} = \varphi^{h,{\rm old}}_{i,j,k} - \frac{\mathcal{L}^h\left(\varphi^{h,{\rm old}}_{i,j,k}\right)}{\frac{\partial\mathcal{L}^h\left(\varphi^{h,{\rm old}}_{i,j,k}\right)}{\partial\varphi_{i,j,k}^{h,{\rm old}}}},\label{phi_new}
\end{equation}
where
\begin{widetext}
\begin{equation}
\frac{\partial\mathcal{L}^h\left(\varphi^{h,{\rm old}}_{i,j,k}\right)}{\partial\varphi_{i,j,k}^{h,{\rm old}}}
\equiv -\frac{6}{h^2} + \frac{\alpha}{|\alpha|} \frac{4(1-3w)}{h^4\sqrt{\alpha^2+4(1-w)\Sigma_{i,j,k}}}\Big(\varphi_{i+1,j,k}+\varphi_{i-1,j,k}+\varphi_{i,j+1,k}+\varphi_{i,j-1,k}+\varphi_{i,j,k+1}+\varphi_{i,j,k-1}-6\varphi_{i,j,k}\Big).
\end{equation}
\end{widetext}
Note that the choice $w=1/3$ also greatly simplifies this expression by making the second term on the right-hand side vanish.

Each time the modified gravity forces are to be updated \citep{Arnold:2019vpg} we initialise the field value in the \textsc{Arepo} AMR grid solver with the solution from the previous timestep. We then perform a number of red-black sweeps to update the field values in the cells according to Eq.~\eqref{phi_new}.
At the end of each iteration, an error is calculated for $\varphi_{i,j,k}$ as,
\begin{equation}\label{eq:error_phi}
    e_{i,j,k} = \frac{r_{i,j,k}}{\alpha}
\end{equation}
where $r_{i,j,k}=\mathcal{L}^h(\hat{\varphi}_{i,j,k})$ (Eq.~\eqref{eq:Lphi}) is the residual for the approximate solution $\hat{\varphi}_{i,j,k}$. 
We stop the iterations when our convergence criterion
\begin{equation}\label{eq:error}
    {\rm max}(e_{i,j,k}) < 10^{-2}\, ,
\end{equation}
is fulfilled. This criterion is equivalent to requesting that the approximate solution for the field in any cell is at least $1\%$ accurate.

%---------------------------------------------------------------
\subsection{Multigrid acceleration}
\label{sec:multigrid}
%---------------------------------------------------------------
To solve the scalar field equation of motion, Eq.~\eqref{eq:phi_code}, we employ the multigrid acceleration technique using V-cycles, following the same prescription as presented by \citet{Arnold:2019vpg}. To numerically solve Eq.~\eqref{eq:operator_eqn}, we start the relaxation with some initial guess of the scalar field, $\varphi_{i,j,k}=0$ in all cells. After a few iterations we have
\begin{equation}\label{eq:Lphi2}
\mathcal{L}^h(\hat{\varphi}^h) = r^h\,,  
\end{equation}
for an approximate solution $\hat{\varphi}^h$ with residual $r^h$. After coarsifying, we obtain the equation on the coarse level,
\begin{equation}\label{eq:Lphi3}
\mathcal{L}^H(\hat{\varphi}^H) = \mathcal{L}(\mathcal{R}\hat{\varphi}^h) - \mathcal{R}r^h\,,
\end{equation}
where $\mathcal{R}$ is the restriction operator which is given by the summation over the 8 daughter cells of the coarse cell. Eq.~\eqref{eq:Lphi3} is used to obtain an approximate coarse-level solution of $\hat{\varphi}^H$. Finally, the fine-level solution can be corrected as,
\begin{equation}\label{eq:Lphi4}
\hat{\varphi}^{h,{\rm new}} =  \hat{\varphi}^h + \mathcal{P}(\hat{\varphi}^H -\mathcal{R}\varphi^h)\,,
\end{equation}
where $\mathcal{P}$ is the prolongation operator. All the finer levels are solved by V-cycles using corrections from the two respectively coarser grid levels.

%---------------------------------------------------------------
\subsection{Force calculation}
\label{sec:force}
%---------------------------------------------------------------
From Eq.~\eqref{eq:poisson_nDGP}, it is straightforward to identify the modified gravity contribution to the gravitational acceleration,
\begin{equation}
 \myvec{a}_{\rm MG} = -\frac{1}{2} \myvec{\nabla} \varphi.
\label{eq:dgp_accel}
\end{equation}
We apply a 5-point finite difference scheme to calculate $\myvec{\nabla} \varphi$ at the centres of cells, and use the cloud-in-cell interpolation (which is the same as the mass assignment scheme to calculate the density field $m(\myvec{x})$) to interpolate the force field from the grid to the particle positions. This method allows us to calculate the fifth force directly from the particle distribution using Eq.~\eqref{eq:dgp_accel}. Recall that the GR force is obtained from {\sc Arepo}'s gravity solver, we only employ the AMR solver to calculate the fifth force.

%--------- Figure --------------
\begin{figure*}
 \centering
 \hspace*{-0.2cm}
\includegraphics[align=t,width=0.47\linewidth]{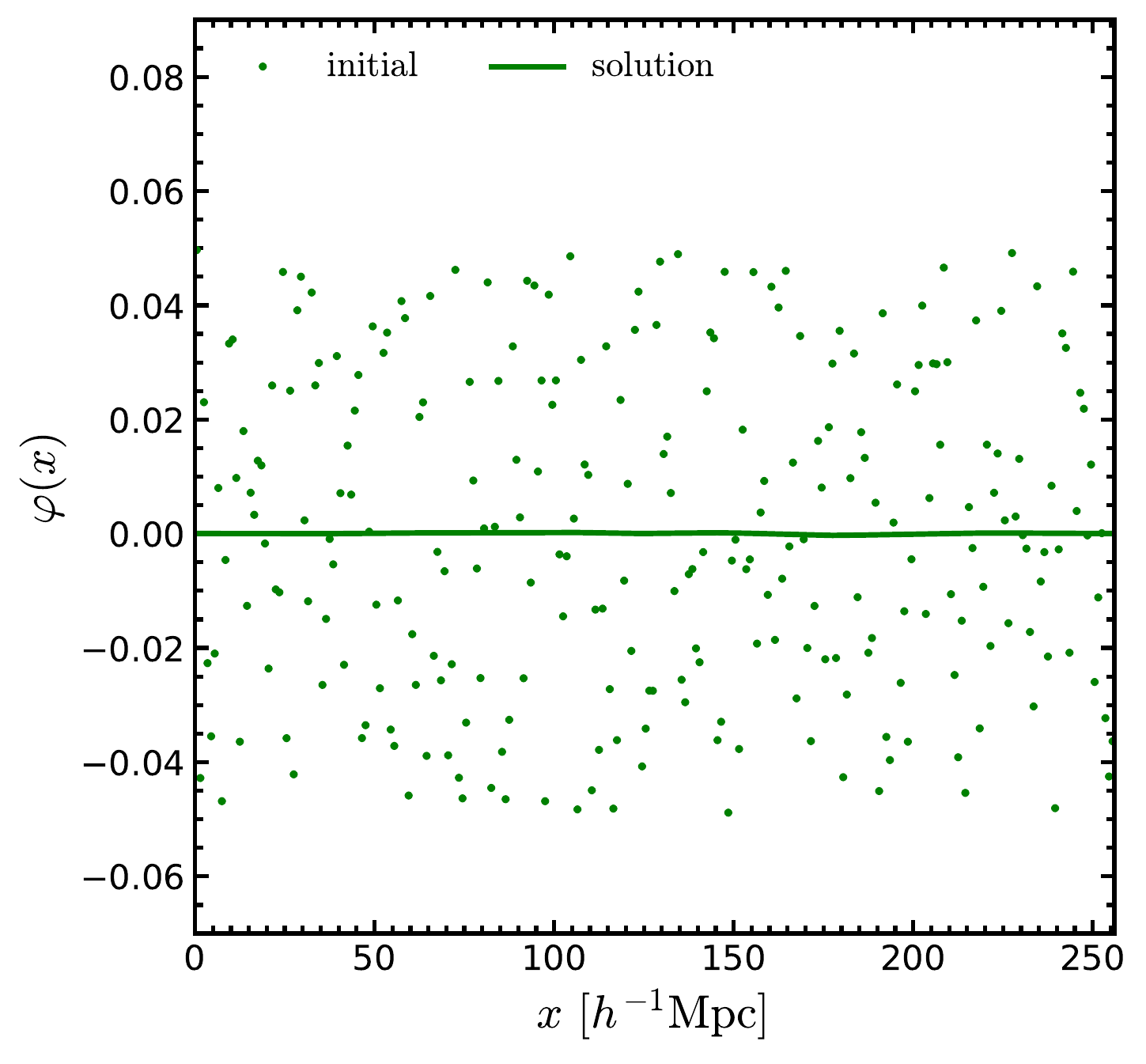}
\includegraphics[align=t,width=0.47\linewidth]{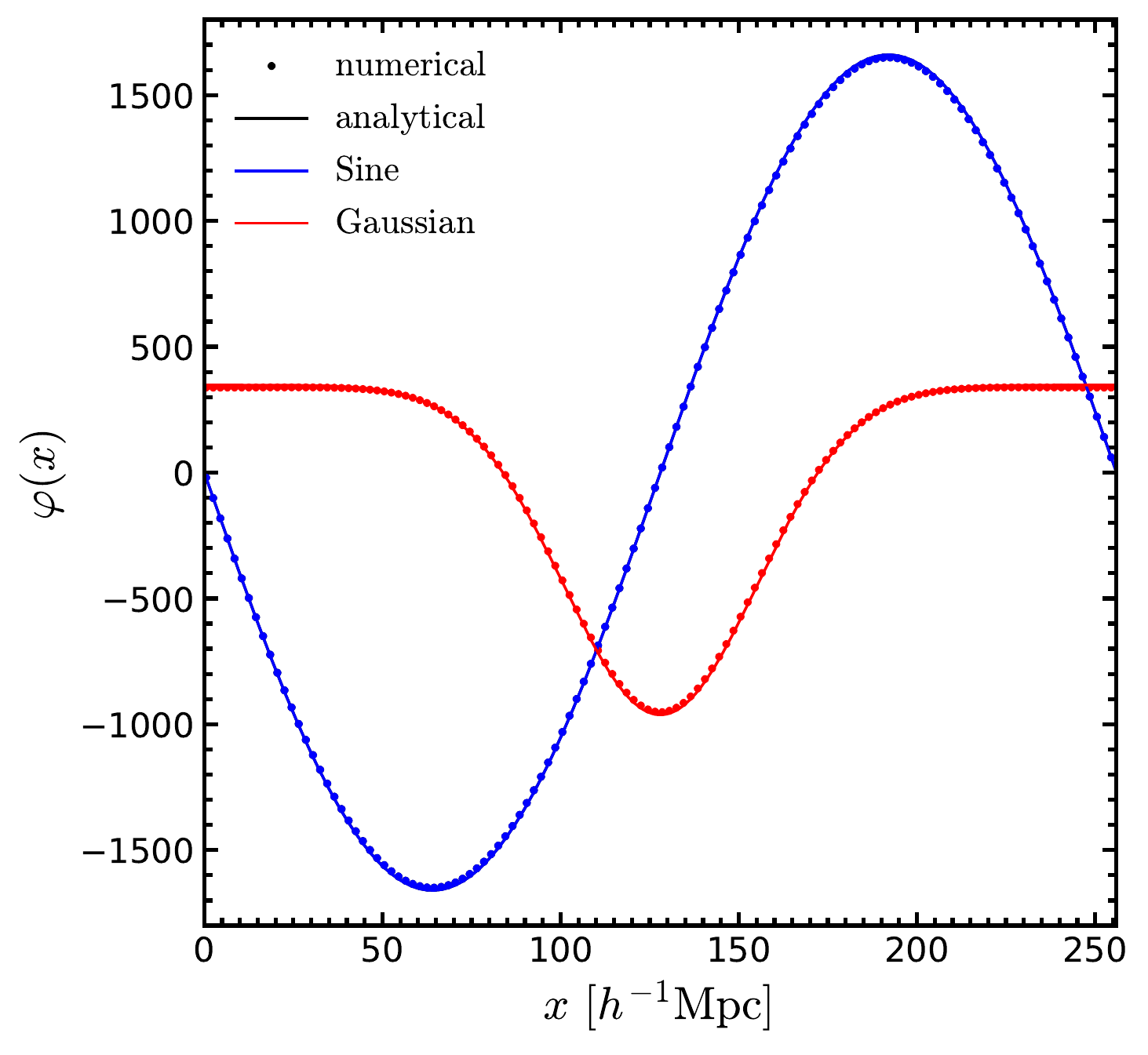}
\hspace*{0.1cm}
\includegraphics[align=t,width=0.45\linewidth]{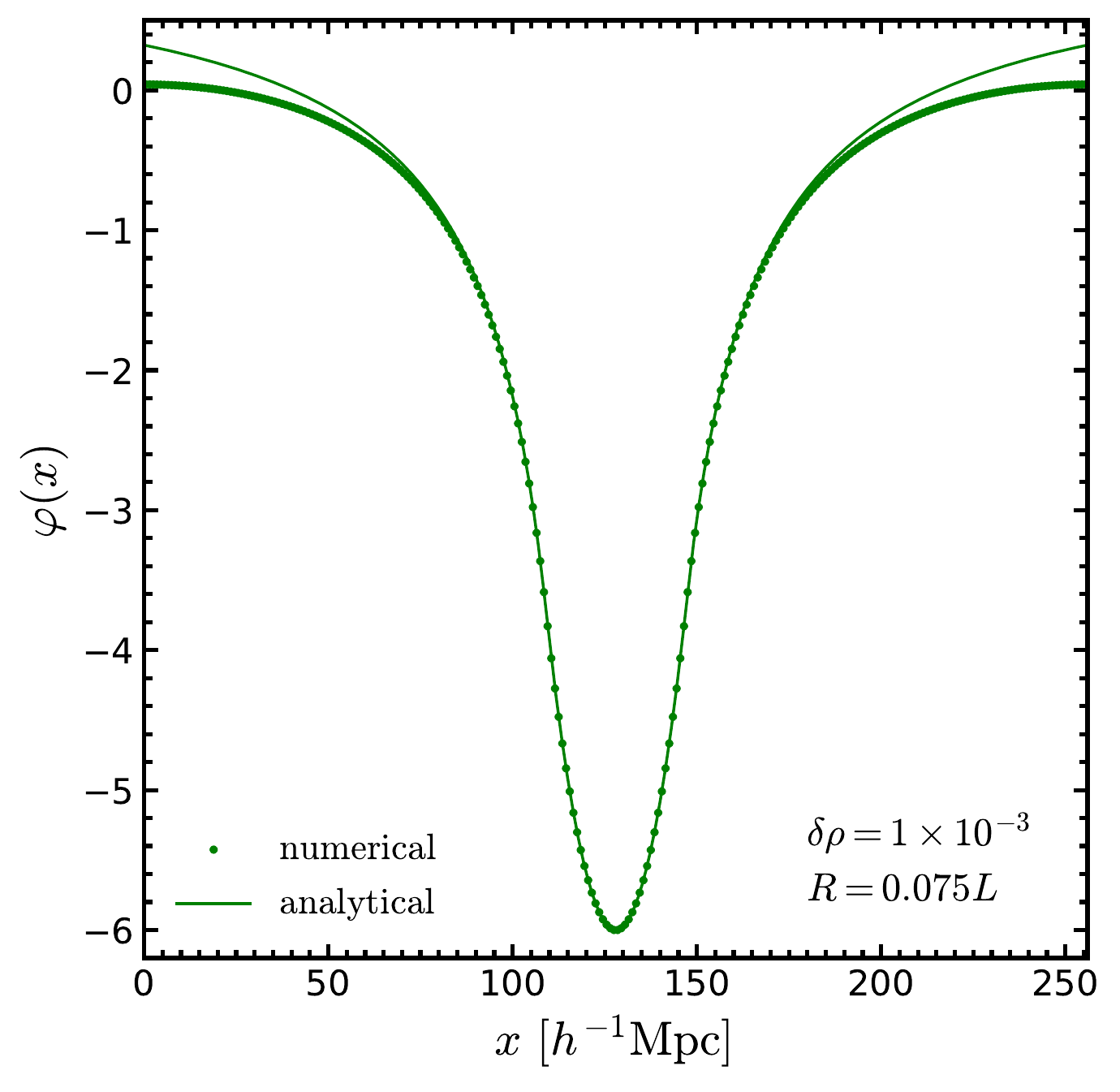}
\hspace*{0.05cm}
\includegraphics[align=t,width=0.47\linewidth]{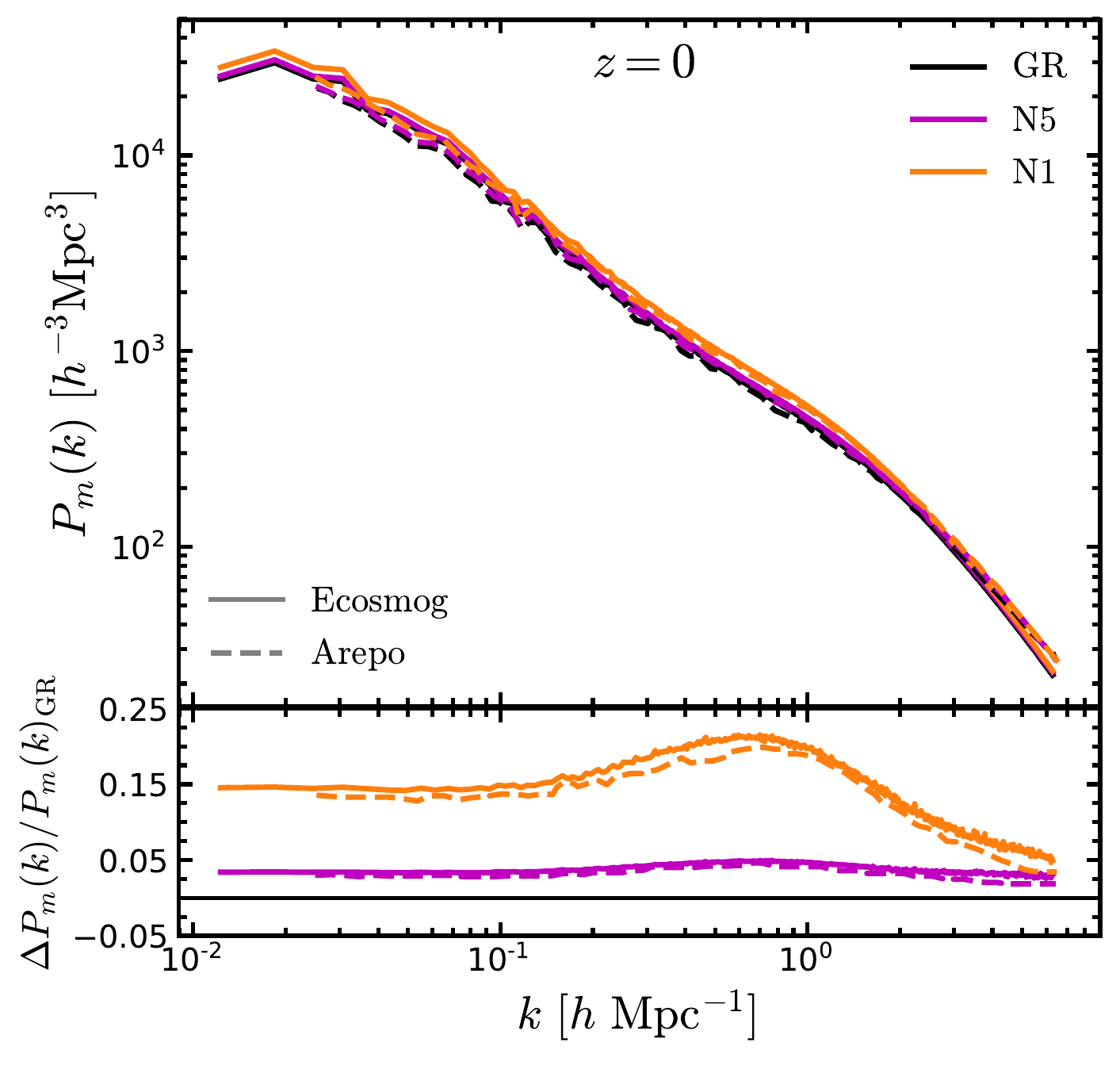}
\caption{Results of code tests. {\it Upper left panel:} Uniform density test, where the green dots represent the random initial values of the scalar field in the range $[-0.05,0.05]$ and the solid green line shows the final solution. {\it Upper right panel:} The 1D sine (blue dots) and Gaussian (red dots) density field tests. The solid lines show the analytical solutions. {\it Lower left panel:} Spherical overdensity test using $\delta\rho = 0.001$ and $R=0.075L$. {The line represents the analytical solution while the dots correspond to our simulation test result.} {\it Lower right panel:} Measured dark matter power spectrum from our test simulations (dashed lines) and from the {\sc elephant} simulations run with the {\sc ecosmog-V} code (solid line). In this panel we show the predictions of GR (black lines), N5 (magenta lines) and N1 (orange lines). The lower subpanel displays the relative difference between the nDGP models and GR for the two different codes.}
\label{fig:code_tests}
\end{figure*}
%---------------------------------------------------------------
\section{Code tests}\label{sec:code_test}
%---------------------------------------------------------------
Following \citet{Li:2013nua}, we perform a series of tests to check that our new {\sc Arepo} MG field solver is working correctly. To do so, we run low-resolution simulations with $256^3$ particles in a box with size $L=256\Mpch$. All tests were performed using the {\sc Arepo} AMR mesh with $2^9$ cells per side at the present time $a=1\,(z=0)$. 

%---------------------------------------------------------------
\subsection{Uniform density field}
%---------------------------------------------------------------
For this test, we have set $\delta\rho_{i,j,k} = 0$ and chosen a set of random values that follow a uniform distribution in the range $[-0.05,0.05]$ as initial guesses of $\varphi_{i,j,k}$. Because the density field is uniform and equal to the cosmological background value, we expect to obtain a smooth and homogeneous $\varphi$. In this test, the residual value, Eq.~\eqref{eq:error}, is reached before the solution converges, for this reason the code stopped when the residual gets a value of ${\rm max}(e_{i,j,k}) < 10^{-6}$, when the solution is well converged. Note that our error criteria is different from that used in {\sc ecosmog-V} \citep{Li:2013nua} (cf. Eq.~\eqref{eq:error_phi}). The result of this test is shown in the upper left panel of Fig.~\ref{fig:code_tests}, where the green dots represent the initial guess and the green solid line is the numerical solution.

%---------------------------------------------------------------
\subsection{One dimensional density field}
%---------------------------------------------------------------
For the first one-dimensional density field test we use a sine-type density field given by,
\begin{equation}\label{eq:rho_sine}
\delta\rho (x) =  \frac{3\beta}{8\pi\,G\,a^2} \sin\left(\frac{2\pi x}{L}\right)\,,
\end{equation}
where $L$ is the box-size. The analytical solution of Eq.~\eqref{eq:phi_code} for this density field is
\begin{equation}\label{eq:phi_sine}
\varphi(x) = -\frac{L^2}{4\pi^2}\sin\left(\frac{2\pi x}{L}\right)\,.
\end{equation}
The solution of this test is presented as blue dots (numerical) and blue solid lines (analytical) in the upper right panel of Fig.~\ref{fig:code_tests}, where we see very good agreement between the two estimates.

The second test uses a Gaussian-type density field, given by
\begin{eqnarray}\label{eq:rho_gauss}
&&\delta\rho(x) = \frac{3\beta}{8\pi\,G\,a^2} \frac{2J\alpha}{w^2}\left[1-2\frac{(x/L-0.5)^2}{w^2}\right]\nonumber\\
&&~~~~~~~~~~~~\times\exp\left[-\frac{(x/L-0.5)^2}{w^2}\right],
\end{eqnarray}
which corresponds to an exact analytic solution
\begin{equation}
    \varphi(x) = L^2J\left[1-\alpha\exp\left(-\frac{(x/L-0.5)^2}{w^2}\right)\right].
\end{equation}
Here $J, \alpha, w$ are constants which we take to be
\begin{equation}
J\ =\ 0.02,~~\alpha\ =\ 0.9999,~~w\ =\ 0.15\,.
\end{equation}
The solution is shown by  the red dots (numerical) and red solid line (analytical) in the upper right panel of Fig.~\ref{fig:code_tests}. Again, the numerical solution agrees very well with the analytic prediction.

%---------------------------------------------------------------
\subsection{Spherical overdensity}
%---------------------------------------------------------------
The previous tests were done using a 1D density field. Now, we test the three dimensional density field. The simplest case is considering the spherically symmetric configuration with constant density.

For the spherical test, it is most convenient to express Eqs.~\eqref{eq:dvarphidr_in}, \eqref{eq:dvarphidr_out} in code units. First of all, since we are assuming that $\delta\rho$ is constant inside the sphere then we can find the expressions for $g_{\rm N}(r)$ and $r_{\rm S}$,
\begin{equation}
g_{\rm N}(r) = \frac{GM(r)}{r^2} = \frac{4\pi G}{3} \delta\rho\,r
\end{equation}
\begin{equation}
r_{\rm S} = \frac{2GM(R)}{c^2} = \frac{8\pi G}{3c^2} \delta\rho\,R^3.
\end{equation}
Hence, using code units and $a=1$, Eqs.~\eqref{eq:dvarphidr_in} and \eqref{eq:dvarphidr_out} can be written as
\begin{equation}
\frac{\rd \varphi}{\rd r} = \frac{3 \beta}{4R_{\rm c}} \left[\sqrt{1+\frac{16 R_{\rm c}}{9\beta^2}\frac{4\pi G}{3} \delta \rho}-1\right]r,\label{eq:phir_in}
\end{equation}
for $r\leq R$ and
\begin{equation}
\frac{\rd {\varphi}}{\rd r} = \frac{3 \beta}{4R_{\rm c}} \left[\sqrt{1+\frac{16R_{\rm c}}{9\beta^2}\frac{4\pi G}{3}\frac{R^3}{r^3} \delta\rho}-1\right]r,\label{eq:phir_out}
\end{equation}
for $r \geq R$, where $r$ is the comoving coordinate, while $R$ is the radius of the spherical overdensity and $\delta\rho$ is the overdensity. 
We place the overdensity in the centre of the simulation box, hence $r$ is given by
\begin{equation}
r = \sqrt{(x - L/2)^2 + (y - L/2)^2 + (z - L/2)^2}\,,
\end{equation}
where $x$, $y$ and $z$ are the Cartesian coordinates. We adopt the values $\delta\rho=0.001, R=0.075L$, and the {\sc Arepo} solution is shown as green dots in the lower-left panel of Fig.~\ref{fig:code_tests}. Meanwhile, given the value of $\varphi(r=0)$, Eqs.~\eqref{eq:phir_in} and \eqref{eq:phir_out} can be integrated to obtain $\varphi(r>0)$ numerically, and the result is shown as the black solid curve in the lower-left panel of Fig.~\ref{fig:code_tests}. 

We can see that the two solutions agree very well, especially at small $r$, i.e., close to the centre of the simulation box, where the overdensity is placed. Far from the centre, the agreement becomes less perfect because the analytical solution does not assume periodicity of the spherical overdensity, while the numerical code uses periodic boundary conditions so that the field sees the overdensities in the replicated boxes as well.

%---------------------------------------------------------------
\subsection{3D matter power spectrum of a cosmological run}
%---------------------------------------------------------------
Finally, we compare the dark matter power spectrum at the present time measured from our test simulations with those from a similar resolution run using the {\sc Ecosmog-V} code \citep{Li:2013nua}, the {\sc elephant} simulations \citep{Paillas:2018wxs}. The lower right panel of Fig.~\ref{fig:code_tests} shows this comparison. We can see that our results display good agreement with previous measurements, and in particular our modified version of {\sc Arepo} reproduces well the amplitude of the power spectrum enhancement in the nDGP model relative to GR on all scales. 

Note that the amplitudes of the matter power spectrum from both codes are slightly different. This is due to the differences in the background cosmology and simulation set-up. The {\sc elephant} simulations were run in a box of size $L=1024\Mpch$ and $N_p=1024^3$ particles using the WMAP-9 simulation parameters \citep{Hinshaw:2012aka}, while the {\sc Arepo} test was run in a box of $256\Mpch$ with $256^3$ dark matter particles using the Planck 15 best-fit parameters \citep{Ade:2015xua}.

%%%%%%%%%%%%%%%%%%%%%%%%%%%%%%%%%%%%%%%%%%%%%%%%%%

% Don't change these lines
\bsp	% typesetting comment
\label{lastpage}
\end{document}